\documentclass[11pt]{article}
	
	\newcommand{\blind}{0} 
	
    \makeatletter
    \renewcommand\section{\@startsection {section}{1}{\z@}%
                                       {-3.5ex \@plus -1ex \@minus -.2ex}%
                                       {2.3ex \@plus.2ex}%
                                       {\normalfont\fontfamily{phv}\fontsize{16}{19}\bfseries}}
    \renewcommand\subsection{\@startsection{subsection}{2}{\z@}%
                                         {-3.25ex\@plus -1ex \@minus -.2ex}%
                                         {1.5ex \@plus .2ex}%
                                         {\normalfont\fontfamily{phv}\fontsize{14}{17}\bfseries}}
    \renewcommand\subsubsection{\@startsection{subsubsection}{3}{\z@}%
                                        {-3.25ex\@plus -1ex \@minus -.2ex}%
                                         {1.5ex \@plus .2ex}%
                                         {\normalfont\normalsize\fontfamily{phv}\fontsize{14}{17}\selectfont}}
    \makeatother
	
 \usepackage{amsmath,amsthm,amsopn,amstext,amsfonts,amssymb,mathrsfs}
	\usepackage{graphicx}
	\usepackage{enumerate}
	\usepackage{natbib} 
	\usepackage{url} 
    \usepackage[dvipsnames]{xcolor}
    \usepackage{comment}
    \usepackage{breqn}
\usepackage{caption}
\usepackage{subcaption}
\usepackage{setspace}
\usepackage[hidelinks]{hyperref}
\usepackage{booktabs}

\def \mca{\mathcal{A}}

\usepackage{bm}

\usepackage{algorithm,algorithmic}
\usepackage{fancybox,float}
\usepackage[dvipsnames]{xcolor}
\usepackage{diagbox}
\usepackage[normalem]{ulem}
\usepackage{booktabs}    
\usepackage{multirow}    
\usepackage{makecell}    
\usepackage{amsmath}     

\begin{document}
		
		\def\spacingset#1{\renewcommand{\baselinestretch}%
			{#1}\small\normalsize} \spacingset{1}
  	
		\if0\blind
		{
			\title{\bf Robust POMDP Framework for Lung Cancer Screening Problems}
			\author{Tong Li $^{a}$, Iakovos Toumazis $^{b}$ and Yisha Xiang $^{a}$ \\
			$^{a}$ Department of Industrial and Systems Engineering, \\
            University of Houston, Houston, TX, USA\\
            $^{b}$ Department of Health Services Research, \\
            Division of Cancer Prevention and Population Sciences, \\
            The University of Texas MD Anderson Cancer Center, Houston, TX, USA}
			\date{}
			\maketitle
		} \fi
		
		\if1\blind
		{

            \title{\bf Robust POMDP Framework for Lung Cancer Screening Problems}
			\author{Author information is purposely removed for double-blind review}
			
\bigskip
			\bigskip
			\bigskip
			\begin{center}
				{\LARGE \bf Robust POMDP Framework for Lung Cancer Screening Problems}
			\end{center}
			\medskip
		} \fi
		\bigskip


	\begin{abstract}

Lung cancer remains a leading cause of cancer mortality, largely because many cases are diagnosed at advanced stages. Low-dose computed tomography (LDCT) screening can reduce lung-cancer mortality by enabling earlier detection. Partially observable Markov decision process (POMDP) models have been increasingly used to personalize screening by maintaining a belief over an individual’s latent cancer state. However, the cancer-state transition probabilities used in these models are often generated from clinical simulations and are subject to estimation error and model misspecification. To address this uncertainty, we propose a robust POMDP framework that accounts for ambiguity in cancer-state transition probabilities through \(\ell_1\)-norm ambiguity sets around the nominal transition probability. The robust model optimizes screening decisions against the worst-case transition probability within these sets, while keeping other model components fixed at their nominal values. Building on the piecewise-linear and convex structure of the robust value function, we adapt a point-based value iteration method to compute robust screening policies. We evaluate the proposed policies using out-of-sample simulations that perturb selected cancer-progression parameters and compare them with the nominal ENGAGE policy for representative female and male heavy-smoker cohorts at age 50. Robust POMDP policies generally outperform nominal ENGAGE in mean out-of-sample quality-adjusted life-years (QALYs), with the best performance achieved at a moderate ambiguity radius within the tested grid. Clinical outcome analysis shows that the robust policy reduces lung cancer deaths (LCDs) in all evaluated settings for the female cohort and in most settings for the male cohort, with additional false positives (FPs). Screening-schedule analysis further shows that the robust policy recommends more LDCT screens and detects more early-stage lung cancers. These findings show that incorporating transition-model uncertainty into data-driven screening models can improve out-of-sample reliability and provide more robust decision support when clinical simulation inputs are misspecified.

	\end{abstract}
			
	\noindent%

\spacingset{1.5} 

\section{Introduction}

Lung cancer is the most frequently diagnosed cancer and the leading cause of cancer death worldwide, with an estimated 2.5 million new cases and 1.8 million deaths in 2022 \citep{bray2024global}. Although age-standardized lung cancer incidence and mortality rates have declined globally in recent decades, lung cancer remains a major public health burden, with substantial variation across regions and populations \citep{Fan2023LungCancerMortality, Li2025GlobalLungCancerTrends}.

Early detection is critical for improving lung cancer outcomes. Randomized lung cancer screening trials, including the National Lung Screening Trial (NLST) and the NELSON trial, have shown that screening high-risk populations can reduce lung-cancer mortality \citep{NLST2011ReducedMortality, deKoning2020nelson}. Because early-stage lung cancer is typically asymptomatic, many cases are not detected until the disease has progressed beyond the localized stage. Survival differs substantially by stage at diagnosis: the 5-year relative survival rate is 65.5\% for localized lung and bronchus cancer, compared with 10.5\% for distant-stage disease \citep{SEERLungCancerStats}. Consequently, detecting lung cancer before metastatic spread is central to reducing lung cancer mortality.

Low-dose computed tomography (LDCT) is currently the most effective method for lung cancer screening, providing detailed information on the size and location of pulmonary nodules. 
In the NLST, LDCT screening reduced lung-cancer mortality compared with chest radiography \citep{NLST2011ReducedMortality}.
Subsequent NLST analyses further show that LDCT detects more early-stage lung cancers than chest radiography \citep{Church2013InitialLDCT, Aberle2013IncidenceScreenings}.
As a result, major health organizations have established evidence-based screening guidelines. 
The U.S. Preventive Services Task Force (USPSTF) currently recommends annual LDCT screening for adults aged 50--80 years who have at least a 20 pack-year smoking history and currently smoke or have quit within the past 15 years \citep{USPSTF2021LungCancerScreening}. Similarly, the American Cancer Society (ACS) recommends yearly LDCT screening for adults aged 50--80 years who currently smoke or formerly smoked and have at least a 20 pack-year smoking history \citep{Wolf2024ACSLungCancerScreening}.


Recent work has proposed risk-based and model-based lung cancer screening approaches that use individual risk factors, life expectancy, smoking history, prior screening results, and downstream outcomes to evaluate screening decisions over time \citep{deKoning2014, Meza2021, Toumazis2021RiskBasedFramework}. These approaches have been formulated using decision-analytic models, cohort or microsimulation models, and POMDP-based sequential decision models \citep{Marshall2001DecisionAnalysis, Manser2005LDCTCostEffectiveness, deKoning2014, Meza2021, Toumazis2021RiskBasedFramework, Petousis2019SequentialDecision}. 
Among these approaches, POMDP-based formulations provide a natural bridge to sequential screening under partial observability.


POMDPs have been increasingly adopted to model sequential medical screening decisions under uncertainty, where a patient’s underlying health states are not directly observable but can be inferred through imperfect screening outcomes \citep{Steimle2021MultiModelMDP}. In particular, the individualized lung cancer screening decision tool (ENGAGE) models the screening process as a POMDP, integrating published lung cancer risk prediction and disease progression models to balance expected health benefits and potential harms over a patient’s lifetime \citep{Toumazis2021RiskBasedFramework}.

However, the performance of POMDP-based screening policies depends strongly on the specification of the disease-progression model, particularly the cancer-state transition probabilities that characterize latent cancer evolution. 
In lung cancer screening models, these probabilities are often generated from clinical simulations, natural-history models, or epidemiologic assumptions. As a result, they are susceptible to estimation error, population heterogeneity, and model misspecification. Policies optimized under a single nominal transition model do not explicitly account for this uncertainty and can perform less reliably when the true cancer-progression dynamics differ from the nominal model.

To enhance the robustness of personalized lung cancer screening decisions under model uncertainty, we extend the POMDP-based lung cancer screening model of \cite{Toumazis2021RiskBasedFramework} and develop a robust partially observable Markov decision process (robust POMDP) framework for lung cancer screening. Our approach accounts for uncertainty in cancer-state transition probabilities by constructing $\ell_1$-norm ambiguity sets around the nominal transition kernel, while keeping the smoking-status transition probabilities, observation probabilities, and reward components fixed at their nominal values. The resulting robust policy is optimized against the worst-case cancer-state transition model within the ambiguity set. In the numerical study, we evaluate the robust policy under perturbed cancer-state transition models and further examine the resulting screening schedules and screen-detected cancers to interpret the observed clinical outcomes.

\section{Literature Review}

\subsection{Lung cancer screening literature}

LDCT screening is currently the primary population-level approach for early lung cancer detection in high-risk groups, but its practical deployment requires balancing mortality reduction against screening-related harms. 
Screening can shift detection toward earlier stages and reduce lung cancer mortality in high-risk populations \citep{NLST2011ReducedMortality}. However, LDCT screening also generates indeterminate findings and false-positive results that prompt repeated imaging, diagnostic work-ups, patient distress, and increased resource use \citep{deKoning2014, Jonas2021}.
These trade-offs motivate research on screening criteria and individualized decision rules that incorporate patient risk profiles, prior screening findings, and downstream diagnostic pathways. 
Comparative modeling studies conducted for the U.S. Preventive Services Task Force evaluate alternative LDCT screening strategies using microsimulation and natural-history models. These studies compare expected benefits and harms, including mortality reduction, life-years gained, false-positive results, overdiagnosis, and screening burden \citep{Meza2021, deKoning2014}.

Decision-analytic modeling has been used to formalize longitudinal lung cancer screening decisions under latent disease status. \cite{Toumazis2021RiskBasedFramework} introduce the ENGAGE tool, a POMDP-based framework that combines published risk prediction and natural-history components with life expectancy and prior screening findings to make individualized screening eligibility decisions over time. Their key methodological contribution is to treat screening as a sequential decision problem in which the patient’s underlying cancer status is not directly observed, and decisions adapt as new screening information becomes available. 
A limitation of this approach, shared by many decision-analytic screening models, is that the cancer-progression transition model is treated as known once specified, even though the natural-history parameters used to construct the cancer-state transition probability can be affected by estimation error, population heterogeneity, and model misspecification.

Data-driven sequential decision methods have also been developed to improve lung cancer screening and diagnostic management. \cite{Petousis2019SequentialDecision} formulate LDCT screening as a learned POMDP using NLST data and show that learned policies can reduce false positives while maintaining favorable cancer detection performance. \cite{Han2023MultipartPOMDP} further advances POMDP-based screening optimization by proposing a multi-part framework to manage different elements of the screening pipeline and control the true-positive versus false-positive trade-off. Related work studies downstream decision making after nodules are detected. \cite{Wang2024DeepRLEarlyDiagnosis} formulate early diagnosis as a sequential decision problem and propose a deep reinforcement learning approach that targets earlier cancer confirmation while controlling unnecessary biopsies and follow-up LDCTs. These studies provide important advances in screening and post-screening decision support, but they generally optimize policies under nominal disease-progression models and do not explicitly hedge against misspecification in cancer-state transition probabilities.

Across these studies, lung cancer screening policies are generally optimized under nominal cancer-progression models that treat the transition probabilities for latent cancer-state evolution, typically estimated from microsimulation or natural-history models, as fixed inputs. In practice, these cancer-state transition probabilities are often generated from clinical simulations or epidemiologic models and may be affected by estimation error and model misspecification.
To address this source of uncertainty, we develop a robust POMDP framework for personalized lung cancer screening that models ambiguity in cancer-state transition probabilities through $\ell_1$-norm ambiguity sets around the nominal transition probability and optimizes screening decisions against the corresponding worst-case transition model. In the proposed formulation, we focus the ambiguity set on the cancer-state transition probability, while specifying the smoking-status transition probabilities, observation probabilities, and reward components according to their nominal estimates from the baseline screening model. This formulation accounts for uncertainty in latent cancer progression while preserving the remaining components of the established modeling framework. To our knowledge, no prior work has optimized lung cancer screening policies using a POMDP formulation that incorporates ambiguity sets for cancer-state transition probabilities.

\subsection{POMDPs in health care applications}

POMDPs provide a natural modeling framework for health care problems in which clinical decisions are made sequentially and the patient’s true disease status is not directly observed. In these settings, clinicians observe imperfect information, such as screening results, symptoms, laboratory findings, or administrative records, and update their beliefs about the underlying health state before making subsequent decisions. \cite{Alagoz2014TutOR} presents a POMDP-based framework for cancer screening design, in which disease states are partially observable and screening tests are imperfect, and shows how the model can be used to optimize screening initiation, cessation, and frequency with long-term objectives such as maximizing total expected quality-adjusted life-years (QALYs).

A large stream of POMDP-based work focuses on cancer screening, where partial observability arises from preclinical disease and noisy diagnostic signals \citep{MaillartIvyRansomDiehl2008,AyerAlagozStoutBurnside2016,SandikciCevikSchacht2018,ErenayAlagozSaid2014}. 
POMDPs have also been used to derive personalized screening and monitoring policies for chronic and infectious diseases using clinical data \citep{KamalzadehAhujaHahslerBowen2021,WuSuen2022,WuCaoSuenLin2024}.
Beyond screening, POMDPs have been applied to other medical decision settings where partial observability is central.  For example, \cite{SuenBrandeauGoldhaberFiebert2018} model the timing of drug sensitivity testing for tuberculosis patients under repeated low-cost but noisy tests versus expensive definitive testing, using a POMDP to balance earlier identification of drug resistance against cost and delayed harms.

These studies show that POMDPs are well suited for clinical decision problems in which decisions depend on latent health states and evolving patient information. However, their policies are typically optimized under specified probabilistic models for disease progression and observation. These model components are often derived from clinical simulations, natural-history models, or epidemiologic assumptions, and the resulting probabilities may be affected by estimation error and model misspecification. Policies optimized under a single nominal model do not explicitly account for such uncertainty, which motivates the robust and distributionally robust POMDP literature reviewed in the next subsection.

\subsection{Robust and distributionally robust POMDPs}

Robust POMDPs extend standard POMDP models by relaxing the assumption that transition and observation probabilities are known exactly. In these models, uncertain transition and/or observation probabilities are typically restricted to a prescribed ambiguity set, and policies are evaluated under adverse models within that set. \cite{ITOH2007453} extend POMDPs to settings in which transition and observation probabilities are imprecisely specified. \cite{osogami15} formulates a robust POMDP in which uncertain parameters are restricted to a given ambiguity set and policies are evaluated against the worst-case model; this work also establishes convexity of the robust value function with respect to the belief state and develops value-iteration methods for robust policy computation. \cite{SAGHAFIAN20181} introduces ambiguous POMDPs to account for ambiguity in the probabilistic model while incorporating the decision maker’s attitude toward ambiguity. Related work also considers robust POMDP formulations motivated by health care settings in which some transition probabilities are difficult to estimate reliably from data \citep{RasouliSaghafian2018, Boloori2020}.

More recent work extends this line of research from parameter ambiguity to distributional ambiguity. 
Rather than considering an ambiguity set directly on a fixed transition or observation probability vector, distributionally robust POMDP models treat the transition-observation probabilities, and in some settings rewards, as random variables generated from an unknown distribution.
\cite{nakao} study a distributionally robust POMDP (DR-POMDP) with moment-based ambiguity sets for the joint distribution of transition-observation probabilities and develop structural and algorithmic results based on the piecewise-linear and convex value-function representation. \cite{li2026distributionally} develop a DR-POMDP framework with distance-based ambiguity sets for uncertain joint distributions of rewards and transition-observation probabilities, derive tractable reformulations for several widely used statistical distances, and establish asymptotic analysis and out-of-sample performance guarantees of optimal policies.

Despite these methodological advances, the use of robust and distributionally robust POMDP methods in individualized lung cancer screening remains limited, particularly for addressing uncertainty in cancer-state transition probabilities derived from microsimulation or natural-history models. The present study adopts a robust POMDP formulation for a clinically motivated screening problem. Specifically, we build on the ENGAGE lung cancer screening POMDP developed by \cite{Toumazis2021RiskBasedFramework}. We construct ambiguity sets around its nominal cancer-state transition probabilities, while keeping the smoking-status transition probabilities, observation probabilities, and reward components fixed at their nominal estimates. This formulation accounts for uncertainty in latent cancer progression and evaluates whether robust screening policies improve out-of-sample performance under misspecified cancer-progression dynamics.

\section{Methods}

\subsection{Problem Description}\label{sec: problem description}
The lung cancer screening problem is a sequential decision process
starting at age 50 and ending at age 100, with decisions made annually.
Each year, an asymptomatic ever-smoker individual, or equivalently, the
healthcare provider acting on the individual's behalf chooses between
two actions: undergo LDCT screening or postpone screening until the next
epoch.

When LDCT screening is performed, the observed result is
classified as either normal or abnormal. In accordance with the clinical
interpretation underlying the Lung CT Screening Reporting and Data System
(Lung-RADS) \citep{acr2014lungrads}, a normal result indicates the
absence of a suspicious finding, whereas an abnormal result indicates the
presence of a suspicious nodule that requires further diagnostic
evaluation to determine whether it is benign or malignant. Following a
normal screening result, the individual continues to the next decision
epoch. By contrast, an abnormal screening result leads to a diagnostic
work-up, which is assumed to be perfectly accurate in our formulation. If
the abnormal finding is ultimately benign, it is treated as a
false-positive result and the individual remains in the decision process.
If lung cancer is confirmed, the individual receives an age- and
stage-specific terminal lump-sum reward representing the expected
postdiagnosis QALYs. The screening process then terminates for that
individual, since patients with diagnosed lung cancer are managed under
a clinical pathway distinct from routine screening candidates.
\citep{Toumazis2021RiskBasedFramework}.

\begin{figure}[t]
\centering

\caption{Lung cancer screening decision process from age \(t\) to age
\(t+1\).}
\label{fig:decision process}

\includegraphics[width=0.8\linewidth]{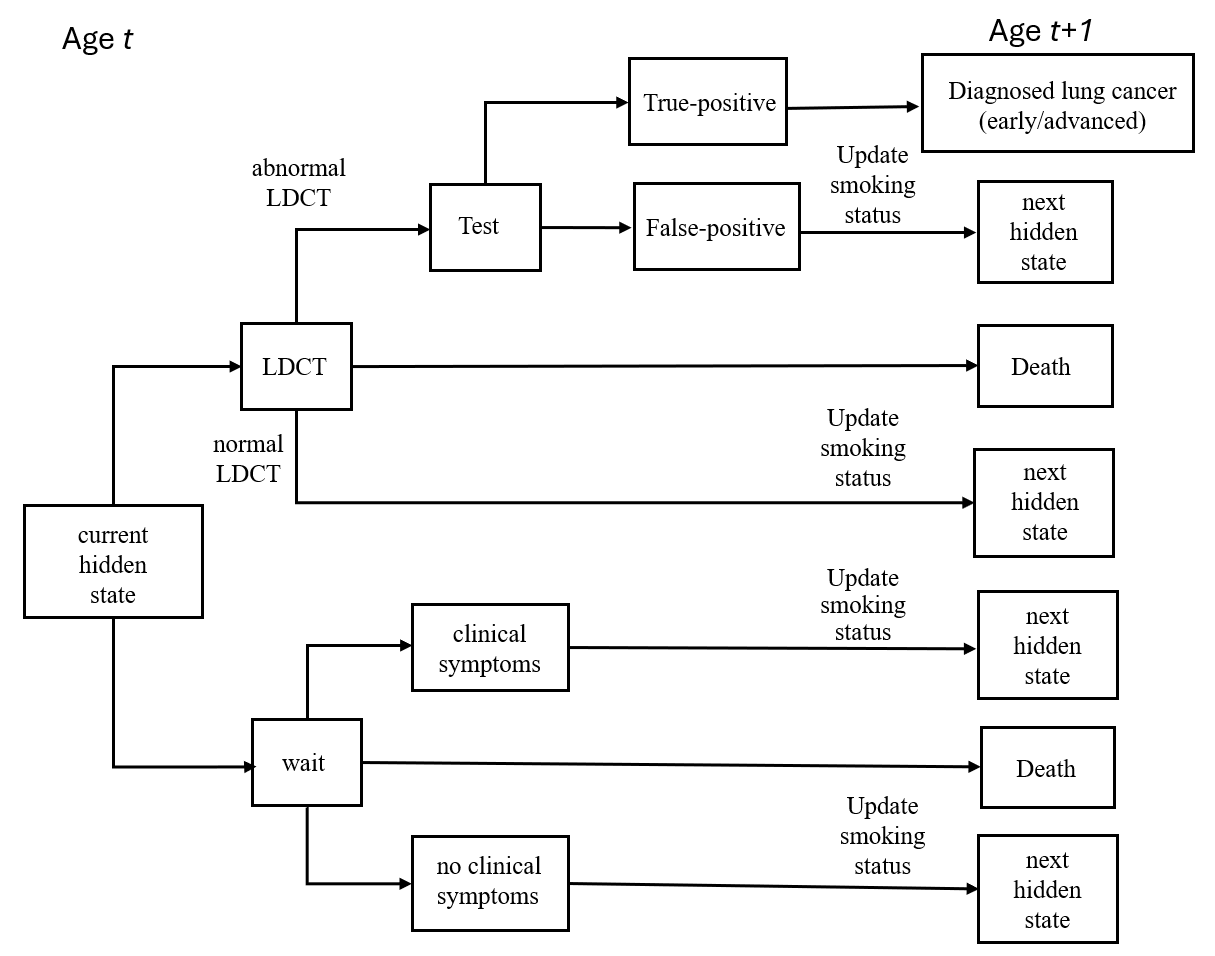}

\vspace{0.5em}

\begin{minipage}{\linewidth}
\small
\textit{Note.} At each annual epoch, the decision maker chooses LDCT screening
or wait, observes the test outcome (abnormal/normal LDCT) or symptom
status (clinical symptoms or none), and the hidden cancer state is
updated for the next epoch. Diagnosed lung cancer (early or advanced) and
death are absorbing states and exit the decision process. The smoking
status is updated before the cancer state and only when the next state
remains partially observable (cancer-free or undiagnosed lung cancer).
Adapted from \citep[Figure~1]{Toumazis2021RiskBasedFramework}.
\end{minipage}

\end{figure}

If screening is deferred at a given decision epoch, the observation in
that period is whether the individual develops a lung cancer--related
clinical symptom, namely hemoptysis \citep{walter2015symptoms}.Between two consecutive decision epochs, the
individual evolves according to the underlying disease and smoking-status
dynamics. Under either action, the individual may die from lung cancer
or from competing causes, in which case the decision process terminates.
Consistent with prior lung cancer screening POMDP models
\citep{Toumazis2021RiskBasedFramework}, all disease-state and
smoking-status transitions are assumed to occur at the end of each
annual decision interval. The overall screening mechanism considered in
this study is illustrated in Figure~\ref{fig:decision process}.

The objective is to determine an optimal screening policy that maximizes
the individual's total expected QALYs over the planning horizon. The reward function adjusts survival time using utility weights that
depend on age, sex, disease stage, and smoking status. 
It also accounts for the
quality-of-life losses associated with LDCT screening, including the
disutility of undergoing screening and the additional disutility induced
by false-positive findings and subsequent diagnostic evaluation.

\subsection{Nominal Lung Cancer Screening POMDP}\label{sec: nominal model engage}
In this section, we briefly introduce the nominal lung cancer screening POMDP in 
\cite{Toumazis2021RiskBasedFramework}, which serves as the baseline clinical decision
model for our study. 
The nominal model is specified by the tuple
\(\bigl(\mathcal{T}, \mathcal{S}, \mathcal{S}^{PO}, \mathcal{M},
\mathcal{A}, \mathcal{O}, \widehat P_t, \widehat Q_t, \widehat Z_t,
R_t, G_t, \beta\bigr)\), whose components are defined as follows.
Decisions are made annually from age 50 until the terminal age
of 100. The set of
decision epochs is denoted by \(\mathcal{T} = \{0, 1, \ldots, T-1\}\),
where epoch \(t \in \mathcal{T}\) corresponds to age \(50 + t\) and
\(T = 50\) is the number of decision epochs in the planning
horizon. 
Future QALYs are discounted by a constant discount factor
\(\beta \in (0,1)\).

The cancer-related state space is
\(\mathcal{S} = \{s_1, s_2, s_3, s_4, s_5, s_6\}\), where \(s_1\)
denotes the cancer-free state, \(s_2\) and \(s_3\) denote undiagnosed
early-stage and undiagnosed advanced-stage lung cancer, \(s_4\) and
\(s_5\) denote diagnosed early-stage and diagnosed advanced-stage lung
cancer, and \(s_6\) denotes death. The diagnosed states \(s_4\) and
\(s_5\), together with the death state \(s_6\), are absorbing terminal
states. The cancer-free and undiagnosed cancer states form the partially
observable subset \(\mathcal{S}^{PO} = \{s_1, s_2, s_3\}\). The
smoking-status state
\(m_t \in \mathcal{M} = \{m_1, m_2, m_3, m_4\}\) represents former smokers and current light, moderate, and heavy smokers, respectively, and is fully observable to the decision maker. The model therefore has a
mixed-observability structure: at each decision epoch \(t\), the
decision maker directly observes \(m_t\) but maintains a belief over the
unobservable cancer state.

At each annual decision epoch, the decision maker chooses an action from
\(\mathcal{A} = \{a_1, a_2\}\), where \(a_1\) denotes LDCT screening
and \(a_2\) denotes waiting until the next decision epoch. The
observation space is \(\mathcal{O} = \{o_1, o_2, o_3, o_4\}\). Under
LDCT screening, \(o_1\) and \(o_2\) denote abnormal and normal screening
results, respectively. Under deferral, \(o_3\) and \(o_4\) denote the
presence and absence of hemoptysis, respectively. 
As shown in Figure~\ref{fig:decision process}, at each epoch \(t\),
the action \(a_t\) is taken first and the observation \(o_t\) is then
realized. If the process does not enter a terminal state during the
decision interval, the smoking status and belief state are updated
before the next decision epoch.

The dynamics of the model are specified by three probability
components. The cancer-state transition probability
\(\widehat P_t(s' \mid s, m, a)\) describes the probability that the
cancer state transitions from \(s\) to \(s'\) between epochs \(t\) and
\(t+1\), conditional on smoking status \(m\) and action \(a\). In our
formulation, \(\widehat P_t\) is action-dependent.\footnote{For
\(a=a_2\) (wait), \(\widehat P_t\) is specified based on the
cancer-state transition structure in
\cite{Toumazis2021RiskBasedFramework}, which characterizes cancer
incidence, preclinical disease progression, and death in the absence of
LDCT screening. For \(a=a_1\) (LDCT screening), \(\widehat P_t\)
additionally includes screening-detected transitions from \(s_2\) to
\(s_4\) and from \(s_3\) to \(s_5\). See
Appendix~\ref{app:cancer-state-P} for details.}
The parameterization of \(\widehat P_t\), including the wait- and
screen-action transition matrices, is provided in
Appendix~\ref{app:cancer-state-P}.
The nominal smoking-status transition probability
\(\widehat Q_t(m' \mid m)\) describes the transition between smoking
states and is parameterized in Appendix~\ref{app: smoking P}. The
nominal observation probability
\(\widehat Z_t(o \mid s, m, a)\) describes the probability of observing
\(o\) conditional on the current cancer state \(s\), smoking status
\(m\), and action \(a\). Its parameterization is provided in
Appendix~\ref{app: observation P}.\footnote{Under the wait action
\(a_2\), the observation probability does not vary with smoking status. We retain \(m\) in the notation \(\widehat Z_t(o \mid s, m, a)\) for
notational consistency with the LDCT observation model.}

Rewards are measured in QALYs. The model
consists of immediate one-period rewards \(R_t(s, m, a, o)\), accrued
during decision epoch \(t\), and lump-sum terminal rewards \(G_t(s)\),
received upon entering an absorbing state \(s \in \{s_4, s_5, s_6\}\)
or upon reaching the terminal age. The dependence of \(R_t\) on \(m\)
captures smoking-status-specific health utilities, while the dependence
on \(o\) captures the disutility associated with realized LDCT screening
outcomes. Following \cite{Toumazis2021RiskBasedFramework}, immediate
rewards are computed using a half-cycle correction. Detailed reward
parameterization, including the construction of \(R_t\) and the
lump-sum terminal reward \(G_t\), is provided in
Appendix~\ref{app: reward}.

\subsection{Robust POMDP Model Formulation}\label{sec: Robust POMDP Model Formulation}

Building on the nominal screening model in
Section~\ref{sec: nominal model engage} and the robustness motivation
in Section~\ref{sec: problem description}, we formulate a robust
counterpart of the nominal POMDP by introducing the uncertainty in the
cancer-state transition probabilities. The smoking-status transition
probabilities, observation probabilities, and reward components are kept
fixed at their nominal values.

To capture the uncertainty in the cancer-state transition probabilities, we construct an \(\ell_1\)-norm ambiguity set centered at
the nominal kernels. For each decision
epoch \(t\), non-absorbing cancer state \(s \in \mathcal{S}^{PO}\), smoking
status \(m \in \mathcal{M}\), and action \(a \in \mathcal{A}\), the
corresponding ambiguity set is defined as
\[
\mathcal{U}_{t, s, m, a}
=
\left\{
P_t(\cdot \mid s, m, a) \in \Delta(\mathcal{S}):
\left\|P_t(\cdot \mid s, m, a) - \widehat P_t(\cdot \mid s, m, a)\right\|_1
\le \epsilon
\right\},
\]
where \(\epsilon \ge 0\) is a common robustness radius applied to all
decision epochs, cancer states, smoking states, and actions. The absorbing-state rows corresponding to $s_4$, $s_5$, and
$s_6$ are kept fixed at the nominal value, since these states are clinically terminal and their
transitions are not subject to model uncertainty.
To avoid clinically infeasible transitions in the cancer-state transitions, any entry of the nominal probability
$\widehat P_t(\cdot \mid s, m, a)$ that equals zero is constrained
to remain zero for every $P_t(\cdot \mid s, m, a) \in
\mathcal{U}_{t, s, m, a}$.
These zero entries correspond to
transitions that are not permitted by the natural-history model: for example, a cancer-free individual cannot transition in a single
year to an advanced or diagnosed cancer state, and no diagnosis can occur under the wait action.


We further assume an $(s, m, a)$-rectangular ambiguity structure
\citep{iyengar, nilim}: the overall ambiguity set for the
cancer-state transition model at epoch $t$ can be expressed as the
Cartesian product
\(
\mathcal{U}_t
=
\bigotimes_{(s, m, a) \in \mathcal{S}^{PO} \times \mathcal{M} \times \mathcal{A}}
\mathcal{U}_{t, s, m, a}.
\)
The rectangularity assumption allows the adversarial transition
distribution to be selected separately for each augmented
state-action triple, yielding a separable structure that makes the
robust dynamic programming formulation computationally tractable \citep{osogami15}.

At each decision epoch \(t\), the patient's true cancer state is not
directly observable prior to diagnosis. The decision maker therefore
maintains a belief state \(b_t \in \Delta(\mathcal{S}^{PO})\) over the
partially observable cancer-state subset, where \(\Delta(\mathcal{X})\)
denotes the probability simplex over a finite set \(\mathcal{X}\). Let
\(b_t(s)\) denote the posterior probability that the patient is in
cancer state \(s \in \mathcal{S}^{PO}\) at decision epoch \(t\),
conditional on the available history of past actions and observations.
The belief vector satisfies
\(\sum_{s \in \mathcal{S}^{PO}} b_t(s) = 1\) and \(b_t(s) \ge 0\) for
all \(s \in \mathcal{S}^{PO}\). 
Together with the fully observable
smoking status \(m_t\), the pair \((b_t, m_t)\) constitutes a
sufficient statistic for decision making under the mixed-observability
structure of the model \citep{pomdp_book}.
The initial belief state at age 50 is constructed from the Bach lung
cancer risk model \citep{Bach2003}, with details provided in
Appendix~\ref{appendix: initial belief state}.

Given a cancer-state transition probability \(P_t \in \mathcal{U}_t \), after action
\(a_t\) is selected and observation \(o_t\) is realized at epoch \(t\),
the belief state is updated according to Bayes' rule
\citep{pomdp_book}.
For each $s' \in \mathcal{S}^{PO}$, the next belief is given by
\begin{equation}\label{eq: belief updated rule}
b_{t+1}(s')
=
f(b_t, m_t, a_t, o_t)(s')
=
\frac{
\sum_{s \in \mathcal{S}^{PO}}
b_t(s)\,
\widehat Z_t(o_t \mid s, m_t, a_t)\,
P_t(s' \mid s, m_t, a_t)
}{
\sum_{\bar s' \in \mathcal{S}^{PO}}
\sum_{s \in \mathcal{S}^{PO}}
b_t(s)\,
\widehat Z_t(o_t \mid s, m_t, a_t)\,
P_t(\bar s' \mid s, m_t, a_t)
}.
\end{equation}
When \(P_t = \widehat P_t\), this update reduces to the nominal
Bayesian belief update.

Let \(\Pi\) denote the set of admissible screening policies. Each policy
\(\pi = \{\pi_t : t \in \mathcal{T}\}\) maps the current belief state
and smoking status to a screening action:
\(a_t = \pi_t(b_t, m_t)\). Since the problem is finite horizon, these
policies are time-dependent. Given a fixed sequence of cancer-state transition
models \(\{P_k\}_{k = t}^{T-1}\) with \(P_k \in \mathcal{U}_k\), the
expected total discounted QALYs of policy \(\pi\) starting from belief
state \(b\) and smoking status \(m\) at epoch \(t\) is defined as
\[
J_t^{\pi}(b, m; \{P_k\})
=
\mathbb{E}^{\pi, \{P_k\}, \widehat Z, \widehat Q}
\left[
\sum_{k=t}^{T-1} \beta^{k-t}
R_k(s_k, m_k, a_k, o_k)
+
\beta^{T-t} G_T(s_T)
\;\middle|\;
b_t = b,\; m_t = m
\right],
\]
where the expectation is evaluated under policy \(\pi\), the
cancer-state transition models \(\{P_k\}_{k=t}^{T-1}\), the nominal observation
probabilities \(\widehat Z\), and the nominal smoking-status transition
probabilities \(\widehat Q\). The terminal reward is specified in Appendix~\ref{app: reward}. If the process enters an absorbing state
\(s_k \in \{s_4, s_5, s_6\}\) before the terminal age, the corresponding
lump-sum reward \(G_k(s_k)\) is received at the time of transition, and
no additional immediate rewards are accrued thereafter. Under this
convention, \(J_t^{\pi}\) represents the total discounted QALYs under
policy \(\pi\), combining immediate QALY rewards accrued during the
screening process with the appropriate lump-sum reward upon absorption or
at the terminal age.

The robust value of policy \(\pi\) is defined as the worst-case
expected total discounted QALYs over all possible cancer-state
transition probabilities \(P_k \in \mathcal{U}_k, k\in \mathcal{T}\):
\begin{equation*} 
V_t^\pi(b, m)
=
\inf_{\{P_k \in \mathcal{U}_k\}_{k=t}^{T-1}}
J_t^{\pi}(b, m; \{P_k\}).
\end{equation*}
The optimal robust screening policy is obtained by maximizing the
robust value over policies in \(\Pi\):
\(V_t(b, m) = \max_{\pi \in \Pi} V_t^\pi(b, m)\).

For each decision epoch \(t \in \mathcal{T}\), belief state $b$ and smoking status $m\in \mathcal{M}$, the optimal robust value function
satisfies the Bellman equation
\begin{equation}\label{eq: Bellman robust pomdp}
\begin{aligned}
V_t(b, m)
=
\max_{a \in \mathcal{A}}
\bigg\{
& \sum_{s \in \mathcal{S}^{PO}} b(s)
\sum_{o \in \mathcal{O}}
\widehat Z_t(o \mid s, m, a)\,
R_t(s, m, a, o) \\[2pt]
& +
\beta
\min_{P_t \in \mathcal{U}_{t}^{a, m}}
\sum_{s \in \mathcal{S}^{PO}} b(s)
\sum_{o \in \mathcal{O}}
\widehat Z_t(o \mid s, m, a)
\sum_{s' \in \mathcal{S}}
P_t(s' \mid s, m, a) \\[2pt]
& \quad\times
\sum_{m' \in \mathcal{M}}
\widehat Q_t(m' \mid m)\,
\mathcal{V}_{t+1}(s', b_{t+1}, m')
\bigg\},
\end{aligned}
\end{equation}
where $b_{t+1} = f(b, m, a, o)$ is the next-epoch belief state, 
computed by the Bayesian update rule \eqref{eq: belief updated rule}, the continuation value is 
\[
\mathcal{V}_{t+1}(s', b_{t+1}, m')
=
\begin{cases}
V_{t+1}(b_{t+1}, m'), & s' \in \mathcal{S}^{PO},\\[4pt]
G_{t+1}(s'), & s' \in \{s_4, s_5, s_6\},
\end{cases}
\]
and
\(
\mathcal{U}_{t}^{a, m}
=
\bigotimes_{s \in \mathcal{S}^{PO}}
\mathcal{U}_{t, s, m, a}
\)
denotes the cancer-state ambiguity set associated with action \(a\)
and smoking status \(m\) at epoch \(t\). 
When \(\epsilon = 0\), each ambiguity set contains only the nominal
transition vector, and the robust POMDP reduces to the nominal lung
cancer screening POMDP described in
Section~\ref{sec: nominal model engage}.

The first term in \eqref{eq: Bellman robust pomdp} represents the
expected immediate QALYs under the nominal observation model. The second term evaluates the discounted continuation value under the worst-case cancer-state transition model in \(\mathcal{U}_{t}^{a,m}\). 
Under the \((s,m,a)\)-rectangular ambiguity structure, the adversarial choice is made over cancer-state transition probabilities indexed by the current
state, smoking status, and action. The
smoking-status transition remains fixed at the nominal probability
\(\widehat Q_t(m' \mid m)\). Thus, the inner minimization of \eqref{eq: Bellman robust pomdp} is over the
cancer-state transition model only, whereas the outer maximization selects the screening action that maximizes worst-case expected QALYs.

The optimal robust value function in \eqref{eq: Bellman robust pomdp} retains the piecewise-linear and convex (PWLC) structure of standard POMDP value functions \citep{osogami15}. Specifically, for each epoch \(t\) and smoking status \(m\), the value function can be represented by the maximum inner product between the
belief state \(b\) and a finite collection of slope vectors \(\Lambda_t^m\) defined over the belief simplex
\(\Delta(\mathcal{S}^{PO})\).
Because smoking status is fully observable, a separate slope-vector collection \(\Lambda_t^m\) is maintained for each \(m \in \mathcal{M}\).
The PWLC structure provides the basis for the point-based value iteration algorithm developed in Section~\ref{sec: robust pomdp solution}. Each point-based backup solves
the inner minimization over \(\mathcal{U}_{t}^{a,m}\) by linear programming.


\subsection{Robust POMDP solution}\label{sec: robust pomdp solution}

We solve the robust POMDP model in
\eqref{eq: Bellman robust pomdp} using a point-based value iteration
algorithm adapted from \cite{osogami15}. Exact solution of a POMDP over
the continuous belief simplex is computationally intractable. In the
robust formulation, this difficulty is further compounded by the inner minimization over transition-probability ambiguity sets. 
We therefore use a point-based approximation in which Bellman equation \eqref{eq: Bellman robust pomdp} are applied at a finite set of representative belief points, and the PWLC representation is used to construct the corresponding slope vectors.

According to the PWLC structure of the robust value function, for each decision epoch
\(t\) and smoking status \(m\), the value function can be written as
\[
V_{t}(b, m)
=
\max_{\alpha^{m} \in \Lambda_{t}^{m}}
\sum_{s \in \mathcal{S}^{PO}} \alpha^{m}(s)\, b(s), \forall b \in \Delta(\mathcal{S}^{PO}).\]
Here, \(\Lambda_t^m\) denotes the slope-vector collection at epoch
\(t\) for smoking status \(m\). The dependence on \(m\) reflects the
mixed-observability structure of the model. Because smoking status is
fully observable, a separate slope-vector collection is maintained for
each \(m \in \mathcal{M}\). This representation allows the value
function to be approximated by dynamic programming over a finite set of representative belief points.

To express the Bellman backup compactly, we aggregate the continuation value over the smoking-status transitions. 
For each observation \(o \in \mathcal{O}\) and each possible next
smoking status \(m' \in \mathcal{M}\), let
\(\alpha_o^{m'} \in \Lambda_{t+1}^{m'}\) denote the slope vector
selected from the next-epoch slope-vector collection associated with
\(m'\). With a slight abuse of notation, the observation dependence is
carried by the superscripted vector \(\alpha_o^{m'}\), while
\(\Lambda_{t+1}^{m'}\) continues to denote the corresponding
next-epoch slope-vector collection. The smoking-aggregated continuation
vector under current smoking status \(m\) is defined as
\(
\bar\alpha_o^{m}(s')
=
\sum_{m' \in \mathcal{M}}
\widehat Q_t(m' \mid m)\,
\psi_{t+1}(s',\alpha_o^{m'}), s' \in \mathcal{S}.
\)
Here, \(\psi_{t+1}\) maps the next cancer state and the selected
slope vector to the corresponding continuation value:
\[
\psi_{t+1}(s',\alpha_o^{m'})
=
\begin{cases}
\alpha_o^{m'}(s'), & s' \in \mathcal{S}^{PO},\\
G_{t+1}(s'), & s' \in \{s_4,s_5,s_6\}.
\end{cases}
\]

For each belief \(b\), smoking status \(m\), and action \(a\), the inner minimization problem of Bellman equation \eqref{eq: Bellman robust pomdp} can be written as the following optimization problems:
\begin{equation}
\label{eq:opt5}
\begin{aligned}
\min_{U_{m},\, P_t} \quad
& \sum_{o \in \mathcal{O}} U_{m}(o) \\
\text{s.t.} \quad
& U_{m}(o) \ge
\sum_{s \in \mathcal{S}^{PO}} b(s)\,
\widehat Z_t(o \mid s, m, a)
\sum_{s' \in \mathcal{S}}
P_t(s' \mid s, m, a)\,
\bar\alpha_o^{m}(s'), \\
&\quad \forall {\alpha}_o^{m'} \in \Lambda_{t+1}^{m'}, \, \forall m' \in \mathcal{M}, \, \forall o \in \mathcal{O}, \\
& P_t(\cdot \mid s, m, a) \in \mathcal{U}_{t,s,m,a},
\qquad \forall s \in \mathcal{S}^{PO}.
\end{aligned}
\end{equation}

Under the \((s,m,a)\)-rectangular ambiguity structure, the adversarial transition vector is selected separately for each current cancer state \(s\), smoking status \(m\), and action \(a\). 
The auxiliary variable \(U_{m}(o)\) represents the upper bound of the expected reward-to-go at observation $o$ and smoking status $m$, evaluated over slope vectors at epoch \(t+1\). 
In an optimal solution of the linear program \eqref{eq:opt5}, a tight constraint involving \(U^{*}_{m}(o)\) identifies a slope vector $\alpha^{*,m}_{o}$ that attains this value.
The same solution also provides the worst-case cancer-state transition probabilities \(P_t^{*}(\cdot \mid s,m,a)\). These quantities provide the inputs for the action-specific robust slope vector $\alpha^{*}_{a}, \forall a \in \mca$.

Based on the $\ell_1$-norm ambiguity set $\mathcal{U}_{t, s, m, a}$ defined in Section~\ref{sec: Robust POMDP Model Formulation}, the optimization problem~\eqref{eq:opt5} can be reformulated as the following linear
program:
\begin{equation*}
\begin{aligned}
\min_{U_{m},\, P_t} \quad
& \sum_{o \in \mathcal{O}} U_{m}(o) \\
\text{s.t.} \quad
& U_{m}(o) \ge
\sum_{s \in \mathcal{S}^{PO}} b(s)\,
\widehat Z_t(o \mid s, m, a)
\sum_{s' \in \mathcal{S}}
P_t(s' \mid s, m, a)\,
\bar\alpha_{o}^{m}(s'), \\
& \quad
\forall \alpha_o^{m'} \in \Lambda_{t+1}^{m'},\, \forall m'\in \mathcal{M}, \, \forall o \in \mathcal{O}, \\[2pt]
& \sum_{s' \in \mathcal{S}} P_t(s' \mid s, m, a) = 1,
\qquad \forall s \in \mathcal{S}^{PO}, 
\end{aligned}
\end{equation*}
\begin{equation}\label{linear program}
\begin{aligned}
& \big\| P_t(\cdot \mid s, m, a) - \widehat P_t(\cdot \mid s, m, a) \big\|_1 \le \epsilon,
\qquad \forall s \in \mathcal{S}^{PO}, \\
& 0 \le P_t(s' \mid s, m, a) \le 1,
\qquad \forall s \in \mathcal{S}^{PO},\ s' \in \mathcal{S}.
\end{aligned}
\end{equation}
The \(\ell_1\)-norm constraint
\(
\|P_t(\cdot \mid s,m,a)-\widehat P_t(\cdot \mid s,m,a)\|_1 \le \epsilon
\)
has an equivalent linear representation obtained by introducing nonnegative auxiliary variables for the element-wise absolute deviations. With this representation, the backup problem \eqref{linear program} becomes a finite-dimensional linear program and can be solved by any off-the-shelf LP solver.

After solving~\eqref{linear program}, let
\(P_t^*(\cdot \mid s,m,a)\) denote the optimal worst-case transition
probabilities for all \(s \in \mathcal{S}^{PO}\). For each observation
\(o\) and smoking status $m$, we identify a slope vector $\alpha^{*,m}_{o}$ achieves the tight
constraint \(U^*(o)\). Therefore, for each \(s \in \mathcal{S}^{PO}\), the robust slope vector associated with action \(a\) is given by
\begin{equation}\label{eq: alpha-vector construction}
\alpha_{a}^{*}(s)
=
\sum_{o \in \mathcal{O}} \widehat Z_t(o \mid s, m, a)\, R_t(s, m, a, o)
+
\beta
\sum_{o \in \mathcal{O}}
\widehat Z_t(o \mid s, m, a)
\sum_{s' \in \mathcal{S}}
P_t^{*}(s' \mid s, m, a)\,
\bar\alpha_{o}^{*}(s'),
\end{equation}
where \(\bar\alpha_o^*(s')\) is the smoking-aggregated continuation
vector evaluated under the maximizing slope-vector 
\(\alpha_o^{*,m}\). For absorbing states
\(s' \in \{s_4,s_5,s_6\}\), the continuation value is the lump-sum
reward \(G_{t+1}(s')\). 
After constructing the action-specific vector
\(\alpha_a^*\) for each \(a \in \mathcal{A}\), the robust value at
belief \(b\) and smoking status \(m\) is computed as
\(
V_t(b,m)
=
\max_{a \in \mathcal{A}}
\sum_{s \in \mathcal{S}^{PO}} b(s)\,\alpha_a^*(s).
\)
The optimal screening action is any action that attains this maximum.

\begin{algorithm}
\caption{Robust Point-Based Value Iteration for Lung Cancer Screening}
\label{alg:robust-pbvi}
\begin{algorithmic}[1]
\REQUIRE Slope vector sets $\{\Lambda_{t+1}^{m'}\}_{m'\in\mathcal{M}}$, belief set $B_0$, smoking status $m$, decision epoch $t$
\ENSURE Updated slope vector set $\Lambda_{t}^{m}$
\STATE $\Lambda_{t}^{m} \leftarrow \emptyset$
\FOR{each belief $b \in B_0$}
    \STATE $\Lambda_{t,b}^{m} \leftarrow \emptyset$
    \FOR{each action $a \in \mathcal{A}$}
        \STATE Solve the linear program~\eqref{linear program} for inputs $(b, m, a)$ and slope vector sets $\{\Lambda_{t+1}^{m'}\}_{m' \in \mathcal{M}}$
        \FOR{each observation $o \in \mathcal{O}$}
            \STATE Extract the maximizer $\alpha_{o}^{*}$ that achieves the optimal $U^{*}(o)$
        \ENDFOR
        \STATE Extract the worst-case transition probability $P_t^{*}(\cdot \mid s, m, a)$ for all $s \in \mathcal{S}^{PO}$ from the optimal solution of~\eqref{linear program}
        \STATE Construct the robust $\alpha$-vector $\alpha_{a}^{*}$ for action $a$ using~\eqref{eq: alpha-vector construction}
        \STATE $\Lambda_{t,b}^{m} \leftarrow \Lambda_{t,b}^{m} \cup \{\alpha_{a}^{*}\}$
    \ENDFOR
    \STATE Select the maximizing slope vector at belief $b$:
    \[
    \Lambda_{t}^{m} \leftarrow
    \Lambda_{t}^{m}
    \cup
    \left\{
        \arg\max_{\alpha \in \Lambda_{t,b}^{m}}
        \sum_{s \in \mathcal{S}^{PO}} b(s)\,\alpha(s)
    \right\}
    \]
\ENDFOR
\RETURN $\Lambda_{t}^{m}$
\end{algorithmic}
\end{algorithm}

Algorithm~\ref{alg:robust-pbvi} summarizes one backward induction step
for a fixed epoch \(t\) and smoking status \(m\). The algorithm follows
the general structure of point-based POMDP methods, but each backup
solves an inner minimization over the transition-probability ambiguity
sets to obtain the worst-case cancer-state transition probabilities.
Because smoking status is fully observable, the algorithm maintains a
separate slope-vector collection \(\Lambda_t^m\) for each
\(m \in \mathcal{M}\). The continuation vectors are aggregated over
next-period smoking status using the nominal transition probabilities
\(\widehat Q_t(m' \mid m)\).

The full backward induction is obtained by applying
Algorithm~\ref{alg:robust-pbvi} for each smoking status
\(m \in \mathcal{M}\) and for each decision epoch \(t\), proceeding
backward from the terminal age to the initial age. The terminal
slope-vector sets are initialized as
\(\Lambda_T^m=\{\alpha_T^m\}\), where
\(\alpha_T^m(s)=G_T(s)\) for all
\(s \in \mathcal{S}^{PO}\) and \(m \in \mathcal{M}\). Here, \(G_T\)
denotes the terminal boundary reward specified in
Appendix~\ref{app: reward}. In the next section, we evaluate the
resulting robust POMDP policy and the nominal ENGAGE policy
\citep{Toumazis2021RiskBasedFramework} under both nominal and perturbed
cancer-state transition models.

\section{Numerical Study}
\label{sec:comp_study}
In this section, we conduct computational studies to evaluate
the performance of the proposed robust POMDP model
\eqref{eq: Bellman robust pomdp}. We perform the analysis for two
representative cohorts: female and male heavy smokers at age 50. 
We evaluate the proposed policies using three groups of outcomes: (i) expected mean out-of-sample QALYs, which capture average performance under uncertainty in cancer-state transition probabilities; (ii) lung cancer deaths (LCDs) and false positives (FPs) per 100{,}000 individuals, which are used to assess the mortality-related benefit and screening-related harm of the resulting policies; and (iii) policy-level screening behavior, including lifetime LDCT screens and screen-detected cancers (SDCs), which helps to characterize the screening mechanisms underlying the clinical outcomes.

\subsection{Numerical Case Setup}
\label{sec:case_setup}

Our model is parameterized using clinically validated data sources and established epidemiologic studies. The key input parameters used in the numerical experiments are summarized in Table~\ref{tab:model_inputs} of Appendix~\ref{app:supplementary-tables}. The discount factor follows the 3\% annual discount rate adopted in \citet{Toumazis2021RiskBasedFramework}, corresponding to $\beta = 1/1.03 \approx 0.97$. Decision epochs are annual, with a planning horizon from age 50 to age 100. Cancer-state transition probabilities are derived from the Bach risk model and the ENGAGE natural-history parameters, smoking-status transition probabilities are obtained from the CISNET smoking history generator, and observation probabilities for LDCT screening and hemoptysis follow values reported in the literature \cite{Toumazis2021RiskBasedFramework}.

The ambiguity set radius $\epsilon$ controls the size of the uncertainty
region around the nominal transition kernel and is varied over 14 values
ranging from $0.10$ to $1.90$. The perturbation intensity $\alpha$ controls
the variance of the mean-corrected log-normal noise applied to the
cancer-progression parameters in the out-of-sample evaluation, and is
varied over seven values from $0.05$ to $3.00$ in order to capture
environments ranging from very mild to severe parameter misspecification.
Section~\ref{sec:impact_radius} describes how the cancer-state transition probabilities are perturbed in the out-of-sample evaluation.

\subsection{Impact of the Ambiguity Set Radius on Patient Outcomes}
\label{sec:impact_radius}

We first investigate how the
radius $\epsilon$ of the ambiguity set affects the out-of-sample performance
of optimal policies derived from robust POMDP models \eqref{eq: Bellman robust pomdp}, in comparison to those from nominal models \citep{Toumazis2021RiskBasedFramework}. We evaluate out-of-sample performance using
the \emph{expected} (mean) total QALYs of an individual, which reflects
average performance under perturbed environments.

Let \(\pi^\ast\) denote the optimal policy obtained from the robust POMDP
model and let \(\pi^\ast_{\mathrm{nom}}\) denote the optimal policy obtained
from the nominal model. We assess out-of-sample performance by applying
both policies to perturbed realizations of the cancer-state transition
kernel. Specifically, we construct a testing set of \(M=300\) Monte Carlo
samples, where each sample corresponds to a perturbed cancer-state
transition kernel \(\widehat P_t^{(i)}\).

The perturbed kernels are generated by applying independent mean-corrected log-normal multipliers to selected cancer-state transition
parameters. The perturbation intensity \(\alpha\) controls the dispersion
of these multipliers, with larger values corresponding to stronger deviations from the nominal cancer-progression model.
In this study, the smoking-status transition
probability \(\widehat Q_t\), the observation probability \(\widehat Z_t\), and the absorbing-state rows are kept fixed at their nominal values, so that the out-of-sample evaluation focuses on misspecification in cancer-progression dynamics. The detailed perturbation procedure is provided in Appendix~\ref{app:perturbation_design}.

For each perturbed kernel $\widehat P_t^{(i)}$, the trained policy is
forward-simulated from the initial belief $b_0$ over the planning horizon
$\mathcal{T}$.  Discounted
QALYs accrued along the simulated trajectory are accumulated to yield
$V^{i}(\pi^\ast)$. 
Because the robust \(\alpha\)-vectors are constructed through
worst-case backups \eqref{eq: alpha-vector construction} over the ambiguity sets \(\mathcal{U}_t\), the resulting policy
accounts for transition uncertainty within these sets. Evaluating
this policy under realized perturbed kernels therefore provides a Monte Carlo assessment of its out-of-sample performance under transition-model
misspecification, consistent with standard practice in the medical
decision-making literature
\citep{Toumazis2021RiskBasedFramework, han2017compliance}.

The expected out-of-sample value of the robust policy is given by
\(
  V^{\pi^\ast} \;=\; \frac{1}{M} \sum_{i=1}^{M} V^{i}(\pi^\ast),
\)
where $V^{i}(\pi^\ast)$ denotes the total expected QALYs obtained by
applying the robust optimal policy $\pi^\ast$ under the $i$-th perturbed
transition kernel $\widehat P_t^{(i)}$. The same evaluation framework is
used to assess the out-of-sample performance of the nominal policy, in
which case $V^{i}(\pi^\ast)$ is replaced by $V^{i}(\pi^\ast_{\mathrm{nom}})$.

We adopt the ENGAGE model \citep{Toumazis2021RiskBasedFramework} with nominal transition and observation
probabilities as the baseline, and compare its optimal lung cancer screening
policies with those derived from the proposed robust POMDP model \eqref{eq: Bellman robust pomdp}.
Specifically, we evaluate and compare the out-of-sample performance of the
optimal policies under 15 values of the ambiguity radius, $\epsilon \in
\{0, 0.10, 0.20, 0.30, 0.40, 0.50, 0.60, 0.70, 0.80, 0.90, 1.00, 1.20, 1.50,$
$1.70, 1.90\}$, while holding the perturbation intensity fixed at
$\alpha = 0.5$. 

\begin{figure}[!htbp]
\centering
\begin{tabular}{cc}
\begin{minipage}{0.5\linewidth}
\centering
\includegraphics[width=\linewidth]{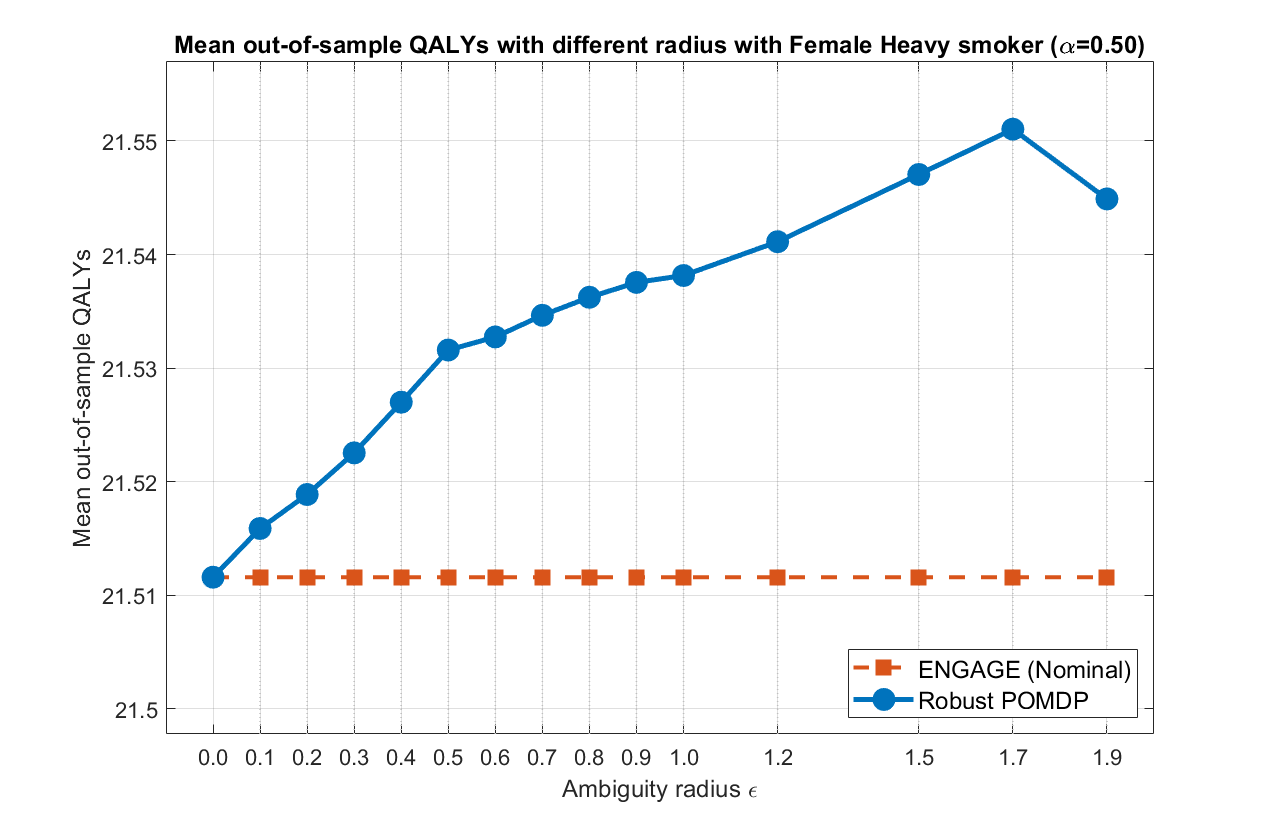}\\
\subcaption*{(a)}
\refstepcounter{subfigure}\label{fig:mean_radius_female}
\end{minipage}
&
\begin{minipage}{0.5\linewidth}
\centering
\includegraphics[width=\linewidth]{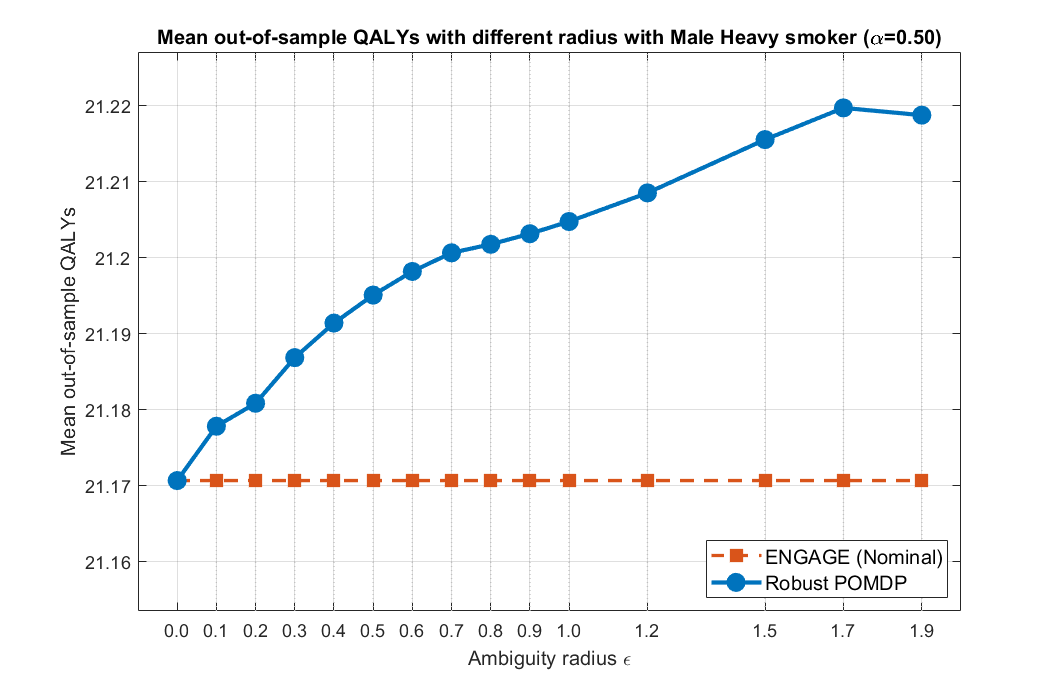}\\
\subcaption*{(b)}
\refstepcounter{subfigure}\label{fig:mean_radius_male}
\end{minipage}
\end{tabular}

\caption{Mean out-of-sample QALYs with ambiguity radius \(\epsilon\) at \(\alpha = 0.5\).
The robust policy matches the ENGAGE baseline at \(\epsilon=0\) and improves for positive ambiguity radii, peaking at \(\epsilon \approx 1.70\).}

\end{figure}

Figures~\ref{fig:mean_radius_female} and \ref{fig:mean_radius_male} report
the mean out-of-sample QALYs for representative female and male heavy smokers, respectively, under different ambiguity-set radii. In each figure, the dashed orange curve corresponds to the nominal ENGAGE policy, while the
solid blue curve corresponds to the proposed robust POMDP policy.

For both cohorts, the robust policy matches the nominal ENGAGE baseline at \(\epsilon=0\) and yields higher mean out-of-sample QALYs for
positive ambiguity radii. The robust policy exhibits an approximately peak-shaped pattern as \(\epsilon\) increases: mean QALYs initially
increase with the ambiguity radius, reach their maximum at around 
\(\epsilon = 1.70\), and then start to decrease. This pattern suggests that accounting for
cancer-state transition uncertainty can improve out-of-sample screening
performance. The magnitude of improvement depends on the
robustness parameter \(\epsilon\). 
When \(\epsilon\) is small, the robust policy yields higher mean
out-of-sample QALYs than the nominal ENGAGE policy, although the improvement remains smaller than that achieved at moderate ambiguity
radii.
When \(\epsilon\) becomes too large, the resulting screening policy becomes overly conservative, and the mean
QALYs begin to decrease from the peak.

For female heavy smokers, the peak mean QALY at
\(\epsilon^\ast = 1.70\) is approximately \(21.5510\), compared with the
nominal value of approximately \(21.5116\), obtaining a gain of
about \(0.0394\) QALYs per individual. For male heavy
smokers, the peak mean QALY is approximately \(21.2197\), yielding a gain of about
\(0.0490\) QALYs per individual. Thus, in these
experiments, \(\epsilon = 1.70\) provides the best observed balance
between robustness and expected out-of-sample performance for both
female and male heavy smokers. The absolute mean-QALY gain for male heavy smokers is about \(24\%\) larger than that for female heavy smokers. 

\subsection{Impact of Perturbation Intensity on Out-of-Sample QALYs}
\label{sec:impact_alpha}


Next, we examine how out-of-sample
performance changes as the perturbation intensity \(\alpha\) varies.
The two parameters play different roles in the experiment. The ambiguity
radius \(\epsilon\) is used when solving the robust POMDP and determines
the size of the transition-probability ambiguity set. In contrast,
\(\alpha\) is used only in the out-of-sample evaluation and controls the
variance of the mean-corrected log-normal perturbations applied to the
nominal cancer-state transition parameters.

As described in Appendix~\ref{app:perturbation_design}, the perturbation
intensity \(\alpha\) controls the dispersion of the mean-corrected log-normal multipliers applied to selected cancer-state transition parameters. Small values of \(\alpha\) generate perturbed transition kernels close to the nominal cancer-progression model, whereas larger
values produce more dispersed out-of-sample environments.

We consider seven perturbation intensities
\(\alpha \in \{0.05,0.10,0.20,0.50,1.00,2.00,3.00\}\). For each
\(\alpha\), we evaluate the nominal ENGAGE policy and the robust POMDP
policies obtained under selected ambiguity radii \(\epsilon\). Under
each \((\alpha,\epsilon)\) combination, both policies are evaluated on
the same set of \(M=300\) perturbed cancer-state transition models, so that the
comparison is based on identical out-of-sample environments. The results
are summarized in Tables~\ref{tab:female_mean_qaly_meancorr} and
\ref{tab:mean_oos_qaly_male} for female and male heavy smokers,
respectively. Each cell reports the ENGAGE baseline value, the robust
POMDP value, the absolute difference ``Diff(R--N)'', and the relative
gain ``RelGain\%'' computed as a percentage of the ENGAGE baseline. The
row \(\epsilon=0\) is included as a reference case: at this radius, the
ambiguity set contains only the nominal transition vector, so the robust
and ENGAGE values are identical.

\begin{table}[h!]
\centering
\caption{Mean out-of-sample QALYs for the Female heavy smoker cohort across selected ambiguity radii \(\epsilon\) and perturbation intensities \(\alpha\).  \label{tab:female_mean_qaly_meancorr}}

\footnotesize
\setlength{\tabcolsep}{4pt}
\begin{tabular}{cl ccccccc}
\toprule
$\epsilon$ & Model & $\alpha=0.05$ & $\alpha=0.10$ & $\alpha=0.20$ & $\alpha=0.50$ & $\alpha=1.00$ & $\alpha=2.00$ & $\alpha=3.00$ \\
\midrule
\multirow{4}{*}{0.00} 
 & ENGAGE     & 21.5099 & 21.5128 & 21.5127 & 21.5116 & 21.5014 & 21.4858 & 21.4826 \\
 & Robust     & 21.5099 & 21.5128 & 21.5127 & 21.5116 & 21.5014 & 21.4858 & 21.4826 \\
 & Diff(R-N)  & $+0.0000$ & $+0.0000$ & $+0.0000$ & $+0.0000$ & $+0.0000$ & $+0.0000$ & $+0.0000$ \\
 & RelGain\%  & $+0.000$  & $+0.000$  & $+0.000$  & $+0.000$  & $+0.000$  & $+0.000$  & $+0.000$  \\
\midrule
\multirow{4}{*}{0.10} 
 & ENGAGE     & 21.5099 & 21.5128 & 21.5127 & 21.5116 & 21.5014 & 21.4858 & 21.4826 \\
 & Robust     & 21.5099 & 21.5128 & 21.5130 & 21.5159 & 21.5060 & 21.4935 & 21.4909 \\
 & Diff(R-N)  & $+0.0000$ & $+0.0000$ & $+0.0003$ & $+0.0043$ & $+0.0046$ & $+0.0078$ & $+0.0082$ \\
 & RelGain\%  & $+0.000$  & $+0.000$  & $+0.001$  & $+0.020$  & $+0.021$  & $+0.036$  & $+0.038$  \\
\midrule
\multirow{4}{*}{0.50} 
 & ENGAGE     & 21.5099 & 21.5128 & 21.5127 & 21.5116 & 21.5014 & 21.4858 & 21.4826 \\
 & Robust     & 21.5300 & 21.5329 & 21.5329 & 21.5316 & 21.5216 & 21.5089 & 21.5109 \\
 & Diff(R-N)  & $+0.0201$ & $+0.0201$ & $+0.0202$ & $+0.0200$ & $+0.0202$ & $+0.0231$ & $+0.0283$ \\
 & RelGain\%  & $+0.094$  & $+0.093$  & $+0.094$  & $+0.093$  & $+0.094$  & $+0.108$  & $+0.132$  \\
\midrule
\multirow{4}{*}{0.90} 
 & ENGAGE     & 21.5099 & 21.5128 & 21.5127 & 21.5116 & 21.5014 & 21.4858 & 21.4826 \\
 & Robust     & 21.5300 & 21.5329 & 21.5332 & 21.5375 & 21.5292 & 21.5209 & 21.5229 \\
 & Diff(R-N)  & $+0.0201$ & $+0.0201$ & $+0.0205$ & $+0.0259$ & $+0.0278$ & $+0.0351$ & $+0.0402$ \\
 & RelGain\%  & $+0.094$  & $+0.093$  & $+0.095$  & $+0.121$  & $+0.129$  & $+0.164$  & $+0.187$  \\
\midrule
\multirow{4}{*}{\textbf{1.70}} 
 & ENGAGE     & 21.5099 & 21.5128 & 21.5127 & 21.5116 & 21.5014 & 21.4858 & 21.4826 \\
 & Robust     & \textbf{21.5496} & \textbf{21.5525} & \textbf{21.5521} & \textbf{21.5510} & \textbf{21.5418} & \textbf{21.5367} & \textbf{21.5411} \\
 & Diff(R-N)  & $\mathbf{+0.0397}$ & $\mathbf{+0.0397}$ & $\mathbf{+0.0394}$ & $\mathbf{+0.0394}$ & $\mathbf{+0.0404}$ & $\mathbf{+0.0510}$ & $\mathbf{+0.0584}$ \\
 & RelGain\%  & $\mathbf{+0.185}$  & $\mathbf{+0.184}$  & $\mathbf{+0.183}$  & $\mathbf{+0.183}$  & $\mathbf{+0.188}$  & $\mathbf{+0.237}$  & $\mathbf{+0.272}$  \\
\midrule
\multirow{4}{*}{1.90} 
 & ENGAGE     & 21.5099 & 21.5128 & 21.5127 & 21.5116 & 21.5014 & 21.4858 & 21.4826 \\
 & Robust     & 21.5494 & 21.5509 & 21.5466 & 21.5449 & 21.5381 & 21.5348 & 21.5386 \\
 & Diff(R-N)  & $+0.0394$ & $+0.0381$ & $+0.0339$ & $+0.0333$ & $+0.0367$ & $+0.0490$ & $+0.0559$ \\
 & RelGain\%  & $+0.183$  & $+0.177$  & $+0.158$  & $+0.155$  & $+0.171$  & $+0.228$  & $+0.260$  \\
\bottomrule
\end{tabular}

\vspace{0.8em}
\begin{minipage}{\linewidth}
\footnotesize
``Diff(R--N)'' denotes the absolute QALY gain of the robust policy over ENGAGE, and ``RelGain\%'' reports this gain as a percentage of the ENGAGE baseline. Bold entries highlight the \(\epsilon=1.70\) row, which attains the largest robust value among the selected radii for each \(\alpha\).
\end{minipage}

\end{table}

\begin{table}[htbp]
\centering
\caption{Mean out-of-sample QALYs for the Male heavy smoker cohort
across selected ambiguity radii \(\epsilon\) and perturbation intensities
\(\alpha\). 
\label{tab:mean_oos_qaly_male}}

\footnotesize
\setlength{\tabcolsep}{4pt}
\begin{tabular}{c l c c c c c c c}
\toprule
$\epsilon$ & Model & $\alpha=0.05$ & $\alpha=0.10$ & $\alpha=0.20$ & $\alpha=0.50$ & $\alpha=1.00$ & $\alpha=2.00$ & $\alpha=3.00$ \\
\midrule
\multirow{4}{*}{0.00}
 & ENGAGE    & 21.1682 & 21.1690 & 21.1681 & 21.1707 & 21.1851 & 21.2210 & 21.2636 \\
 & Robust    & 21.1682 & 21.1690 & 21.1681 & 21.1707 & 21.1851 & 21.2210 & 21.2636 \\
 & Diff(R-N) & +0.0000 & +0.0000 & +0.0000 & +0.0000 & +0.0000 & +0.0000 & +0.0000 \\
 & RelGain\% & +0.000  & +0.000  & +0.000  & +0.000  & +0.000  & +0.000  & +0.000  \\
\midrule
\multirow{4}{*}{0.10}
 & ENGAGE    & 21.1682 & 21.1690 & 21.1681 & 21.1707 & 21.1851 & 21.2210 & 21.2636 \\
 & Robust    & 21.1682 & 21.1690 & 21.1684 & 21.1778 & 21.1909 & 21.2285 & 21.2725 \\
 & Diff(R-N) & +0.0000 & +0.0000 & +0.0002 & +0.0072 & +0.0058 & +0.0076 & +0.0089 \\
 & RelGain\% & +0.000  & +0.000  & +0.001  & +0.034  & +0.027  & +0.036  & +0.042  \\
\midrule
\multirow{4}{*}{0.50}
 & ENGAGE    & 21.1682 & 21.1690 & 21.1681 & 21.1707 & 21.1851 & 21.2210 & 21.2636 \\
 & Robust    & 21.1903 & 21.1910 & 21.1902 & 21.1951 & 21.2064 & 21.2430 & 21.2892 \\
 & Diff(R-N) & +0.0221 & +0.0220 & +0.0220 & +0.0244 & +0.0213 & +0.0220 & +0.0256 \\
 & RelGain\% & +0.104  & +0.104  & +0.104  & +0.115  & +0.100  & +0.104  & +0.121  \\
\midrule
\multirow{4}{*}{0.90}
 & ENGAGE    & 21.1682 & 21.1690 & 21.1681 & 21.1707 & 21.1851 & 21.2210 & 21.2636 \\
 & Robust    & 21.1903 & 21.1913 & 21.1934 & 21.2032 & 21.2155 & 21.2526 & 21.2990 \\
 & Diff(R-N) & +0.0221 & +0.0223 & +0.0253 & +0.0325 & +0.0303 & +0.0317 & +0.0354 \\
 & RelGain\% & +0.104  & +0.105  & +0.120  & +0.153  & +0.143  & +0.149  & +0.166  \\
\midrule
\multirow{4}{*}{\textbf{1.70}}
 & ENGAGE    & 21.1682 & 21.1690 & 21.1681 & 21.1707 & 21.1851 & 21.2210 & 21.2636 \\
 & Robust    & \textbf{21.2158} & \textbf{21.2165} & \textbf{21.2154} & \textbf{21.2197} & \textbf{21.2270} & \textbf{21.2641} & \textbf{21.3118} \\
 & Diff(R-N) & \textbf{+0.0475} & \textbf{+0.0475} & \textbf{+0.0473} & \textbf{+0.0490} & \textbf{+0.0419} & \textbf{+0.0431} & \textbf{+0.0482} \\
 & RelGain\% & \textbf{+0.225}  & \textbf{+0.224}  & \textbf{+0.224}  & \textbf{+0.232}  & \textbf{+0.198}  & \textbf{+0.203}  & \textbf{+0.227}  \\
\midrule
\multirow{4}{*}{1.90}
 & ENGAGE    & 21.1682 & 21.1690 & 21.1681 & 21.1707 & 21.1851 & 21.2210 & 21.2636 \\
 & Robust    & 21.2158 & 21.2165 & 21.2154 & 21.2188 & 21.2264 & 21.2636 & 21.3116 \\
 & Diff(R-N) & +0.0475 & +0.0475 & +0.0473 & +0.0481 & +0.0412 & +0.0426 & +0.0480 \\
 & RelGain\% & +0.225  & +0.224  & +0.224  & +0.227  & +0.195  & +0.201  & +0.226  \\
\bottomrule
\end{tabular}


\end{table}


Two observations emerge from Tables~\ref{tab:female_mean_qaly_meancorr}
and \ref{tab:mean_oos_qaly_male}. First, for most positive ambiguity
radii, the advantage of the robust policy over the nominal ENGAGE baseline tends to increase as the perturbation intensity \(\alpha\) increases, although the pattern is not strictly monotone. This trend suggests that the robust POMDP provides greater out-of-sample benefit when the evaluation
environment is generated with larger perturbation variance. 
Second, $\epsilon=1.70$
provides the best or near-best observed performance among the selected ambiguity radii for both cohorts. This suggests that a moderate robustness radius provides a favorable balance between nominal
performance and protection against cancer-state transition-model misspecification.

For female heavy smokers, the benefit of the robust policy is larger at
higher perturbation intensities, especially at larger ambiguity radii,
although the pattern is not strictly monotone. At $\epsilon=1.70$, the
absolute QALY gain increases from $+0.0397$ at $\alpha=0.05$ to
$+0.0584$ at $\alpha=3.00$, with the corresponding relative gain
increasing from $+0.185\%$ to $+0.272\%$. This indicates that, for female
heavy smokers, stronger perturbations are associated with larger
out-of-sample gains from the robust policy at the best-performing
ambiguity radius.

For male heavy smokers, the robust policy also generally outperforms the
ENGAGE baseline across the selected positive ambiguity radii. However,
unlike the female cohort, the gain does not show a clear increasing
trend as $\alpha$ becomes larger. At $\epsilon=1.70$, the absolute QALY
gain ranges from $+0.0419$ to $+0.0490$ across the tested perturbation
intensities, and the relative gain ranges from $+0.198\%$ to $+0.232\%$.
These results suggest that, for male heavy smokers, the robust policy
provides a stable out-of-sample improvement over the ENGAGE baseline,
although the improvement does not increase consistently with the
perturbation intensity.

\subsection{Lung Cancer Deaths versus False Positives}
\label{sec:lcd_fp}

To complement the QALY-based evaluation, we examine two clinically
interpretable outcomes commonly reported in the lung cancer screening
literature: lung cancer deaths (LCDs) and false positives (FPs) per
100{,}000 individuals over the planning horizon
\citep{Toumazis2021RiskBasedFramework}. These outcomes help characterize
how the robust policy changes mortality-related benefits and
screening-related harms relative to the nominal ENGAGE policy.

LCDs are accumulated through two routes. First, when an individual
reaches a diagnosed cancer state, \(s_4\) or \(s_5\), lung cancer mortality is 
attributed using SEER 5-year survival rates \citep{seer2023}. 
Second, when an individual transitions from an undiagnosed cancer state
\((s_2\) or \(s_3)\) directly into the aggregate death state \(s_6\), we
decompose the death outcome for accounting purposes into lung-cancer
death \(s_6^{\mathrm{LC}}\) and competing-cause death
\(s_6^{\mathrm{other}}\) by using the age-specific Bach mortality $\mu_t$ 
\citep{Bach2003}. 
An FP is counted
whenever an individual in the cancer-free state \(s_1\) undergoes LDCT
screening and receives an abnormal screening observation \(o_1\).

For each perturbation intensity \(\alpha\) and ambiguity radius
\(\epsilon\), we estimate LCDs and FPs through Monte Carlo simulation.
We simulate \(N=100{,}000\) individuals from age 50 to age 100 under both the ENGAGE policy \(\pi^{*}_{\mathrm{nominal}}\) and the
robust policy \(\pi^{*}\), and record the LCD and FP counts in each simulated cohort. To quantify the value of robustness, we compare the robust policy with
the nominal ENGAGE policy for each fixed \((\alpha,\epsilon)\)
setting. Both policies are evaluated under the same perturbed transition
kernels, so the paired differences measure the incremental effect of the
robust policy relative to the nominal one under the same realized environment.

To compare the robust policy with the ENGAGE baseline, we compute paired
differences
\(
\Delta\mathrm{LCD}_{j}
=
\mathrm{LCD}_{j}^{\mathrm{robust}}
-
\mathrm{LCD}_{j}^{\mathrm{nominal}}\) and
\(
\Delta\mathrm{FP}_{j}
=
\mathrm{FP}_{j}^{\mathrm{robust}}
-
\mathrm{FP}_{j}^{\mathrm{nominal}}.
\)
A negative value of \(\Delta\mathrm{LCD}\) indicates that the robust policy produces fewer lung cancer deaths than ENGAGE, whereas a positive
value of \(\Delta\mathrm{FP}\) indicates that the robust policy produces more false positives. We report the sample means of these paired
differences across the \(M=300\) perturbation realizations, together
with their 95\% confidence intervals. We focus on three perturbation
intensities, \(\alpha \in \{0.5,1.0,3.0\}\), ranging from moderate to
severe perturbation levels, and five ambiguity radii,
\(\epsilon \in \{0.10,0.50,0.90,1.70,1.90\}\). The results are reported
in Tables~\ref{tab:lcdfp_female_seer05} and
\ref{tab:lcdfp_male_seer05} for female and male heavy smokers,
respectively. 
All results in the following tables are rounded to the nearest integer.

%
%

\begin{table}[h]
\centering
\caption{LCDs and FPs per 100{,}000 Female Heavy smokers under ENGAGE and Robust POMDP, across ambiguity radii $\epsilon$ and perturbation intensities $\alpha$. 
\label{tab:lcdfp_female_seer05}}

\footnotesize
\begin{tabular}{cccccc}
\toprule
$\alpha$ & $\epsilon$ & LCD per 100k & FP per 100k & $\Delta$LCD & $\Delta$FP \\
         &            & Robust       & Robust      & (R $-$ N)   & (R $-$ N) \\
\midrule
\multirow{6}{*}{0.5} & ENGAGE & \textbf{3332~[3297, 3366]} & \textbf{212807~[212488, 213127]} & --- & --- \\
 & 0.10 & 3327~[3293, 3361] & 212857~[212551, 213163] & $-5~[-14, +4]$ & $+50~[-62, +161]$ \\
 & 0.50 & 3324~[3290, 3359] & 212931~[212624, 213238] & $-8~[-16, +2]$ & $+124~[+13, +234]$$^{*}$ \\
 & 0.90 & 3320~[3286, 3355] & 212894~[212583, 213205] & $-11~[-20, -2]$$^{*}$ & $+86~[-23, +195]$ \\
 & 1.70 & 3322~[3288, 3357] & 212888~[212577, 213198] & $-10~[-19, +0]$$^{*}$ & $+80~[-23, +184]$ \\
 & 1.90 & 3325~[3290, 3359] & 212863~[212557, 213169] & $-7~[-16, +2]$ & $+56~[-52, +164]$ \\
\midrule
\multirow{6}{*}{1.0} & ENGAGE & \textbf{3320~[3254, 3387]} & \textbf{211809~[211065, 212553]} & --- & --- \\
 & 0.10 & 3315~[3248, 3383] & 212989~[212382, 213596] & $-5~[-14, +4]$ & $+1180~[+852, +1509]$$^{*}$ \\
 & 0.50 & 3312~[3245, 3379] & 213014~[212404, 213624] & $-8~[-17, +1]$ & $+1205~[+881, +1530]$$^{*}$ \\
 & 0.90 & 3311~[3244, 3377] & 213040~[212428, 213651] & $-10~[-18, -1]$$^{**}$ & $+1231~[+905, +1557]$$^{**}$ \\
 & 1.70 & 3308~[3241, 3374] & 213063~[212448, 213679] & $-13~[-22, -4]$$^{**}$ & $+1254~[+927, +1582]$$^{**}$ \\
 & 1.90 & 3304~[3237, 3372] & 213049~[212433, 213665] & $-16~[-25, -7]$$^{**}$ & $+1240~[+910, +1571]$$^{**}$ \\
\midrule
\multirow{6}{*}{3.0} & ENGAGE & \textbf{3205~[3028, 3382]} & \textbf{210913~[208977, 212849]} & --- & --- \\
 & 0.10 & 3201~[3024, 3379] & 212137~[210248, 214027] & $-4~[-13, +6]$ & $+1224~[+784, +1664]$$^{*}$ \\
 & 0.50 & 3177~[3000, 3354] & 214154~[212313, 215995] & $-28~[-39, -17]$$^{**}$ & $+3241~[+2749, +3734]$$^{**}$ \\
 & 0.90 & 3173~[2995, 3351] & 214869~[213033, 216705] & $-32~[-44, -20]$$^{**}$ & $+3956~[+3476, +4436]$$^{**}$ \\
 & 1.70 & 3167~[2989, 3345] & 214879~[213046, 216713] & $-38~[-50, -26]$$^{**}$ & $+3966~[+3470, +4462]$$^{**}$ \\
 & 1.90 & 3169~[2992, 3346] & 214878~[213047, 216710] & $-36~[-49, -23]$$^{**}$ & $+3966~[+3471, +4460]$$^{**}$ \\
\bottomrule
\end{tabular}

\vspace{0.8em}

\begin{minipage}{\linewidth}
\footnotesize
Values are mean with 95\% confidence interval over $M = 300$ perturbation samples, each simulating $N = 100{,}000$ individuals. Paired differences $\Delta$LCD and $\Delta$FP compare Robust vs.\ ENGAGE under identical perturbations. Significance: $^{*}$ denotes 95\% CI excluding zero; $^{**}$ denotes both LCD and FP differences significant.
\end{minipage}

\end{table}

\begin{table}[h]
\centering
\caption{LCDs and FPs per 100{,}000 Male Heavy smokers under ENGAGE and Robust POMDP, across ambiguity radii $\epsilon$ and perturbation intensities $\alpha$. 
\label{tab:lcdfp_male_seer05}}

\footnotesize
\begin{tabular}{cccccc}
\toprule
$\alpha$ & $\epsilon$ & LCD per 100k & FP per 100k & $\Delta$LCD & $\Delta$FP \\
         &            & Robust       & Robust      & (R $-$ N)   & (R $-$ N) \\
\midrule
\multirow{6}{*}{0.5} & ENGAGE & \textbf{2499~[2473, 2525]} & \textbf{187012~[186510, 187514]} & --- & --- \\
 & 0.10 & 2508~[2482, 2534] & 185039~[184752, 185325] & $+9~[+1, +17]$$^{**}$ & $-1973~[-2354, -1592]$$^{**}$ \\
 & 0.50 & 2494~[2468, 2519] & 192666~[192376, 192956] & $-5~[-14, +3]$ & $+5654~[+5271, +6037]$$^{*}$ \\
 & 0.90 & 2489~[2463, 2515] & 192679~[192391, 192967] & $-10~[-18, -2]$$^{**}$ & $+5667~[+5287, +6046]$$^{**}$ \\
 & 1.70 & 2487~[2462, 2513] & 192679~[192385, 192973] & $-12~[-20, -3]$$^{**}$ & $+5667~[+5282, +6052]$$^{**}$ \\
 & 1.90 & 2490~[2464, 2516] & 192594~[192297, 192890] & $-9~[-16, -1]$$^{**}$ & $+5582~[+5199, +5965]$$^{**}$ \\
\midrule
\multirow{6}{*}{1.0} & ENGAGE & \textbf{2497~[2447, 2548]} & \textbf{187818~[187069, 188567]} & --- & --- \\
 & 0.10 & 2504~[2453, 2555] & 185206~[184638, 185775] & $+7~[-2, +15]$ & $-2612~[-3027, -2196]$$^{*}$ \\
 & 0.50 & 2483~[2432, 2533] & 192749~[192176, 193322] & $-15~[-23, -7]$$^{**}$ & $+4931~[+4510, +5352]$$^{**}$ \\
 & 0.90 & 2482~[2430, 2534] & 192774~[192204, 193344] & $-16~[-23, -8]$$^{**}$ & $+4956~[+4539, +5373]$$^{**}$ \\
 & 1.70 & 2487~[2436, 2539] & 191898~[191249, 192548] & $-10~[-18, -2]$$^{**}$ & $+4080~[+3642, +4519]$$^{**}$ \\
 & 1.90 & 2485~[2433, 2536] & 191714~[191062, 192365] & $-12~[-21, -4]$$^{**}$ & $+3896~[+3456, +4335]$$^{**}$ \\
\midrule
\multirow{6}{*}{3.0} & ENGAGE & \textbf{2464~[2321, 2607]} & \textbf{188300~[186462, 190138]} & --- & --- \\
 & 0.10 & 2472~[2330, 2614] & 184861~[183159, 186562] & $+8~[-1, +16]$ & $-3439~[-3973, -2906]$$^{*}$ \\
 & 0.50 & 2441~[2299, 2584] & 192727~[191002, 194453] & $-23~[-31, -15]$$^{**}$ & $+4428~[+3856, +4999]$$^{**}$ \\
 & 0.90 & 2433~[2291, 2576] & 193727~[191924, 195530] & $-31~[-40, -22]$$^{**}$ & $+5427~[+4774, +6080]$$^{**}$ \\
 & 1.70 & 2438~[2296, 2579] & 190801~[188994, 192608] & $-26~[-36, -17]$$^{**}$ & $+2501~[+1937, +3066]$$^{**}$ \\
 & 1.90 & 2438~[2297, 2580] & 190713~[188917, 192510] & $-26~[-35, -16]$$^{**}$ & $+2414~[+1852, +2975]$$^{**}$ \\
\bottomrule
\end{tabular}

\end{table}

Tables~\ref{tab:lcdfp_female_seer05} and
\ref{tab:lcdfp_male_seer05} report paired differences between the robust
policy and the nominal ENGAGE policy in LCDs and FPs under the same
perturbed evaluation environments. Across all evaluated settings for the
female cohort and most evaluated settings for the male cohort, the
robust policy produces fewer LCDs than the nominal ENGAGE policy. In all
settings where LCDs are reduced, FPs increase, showing that the mortality
benefit of the robust policy is accompanied by additional false-positive
findings.

This pattern suggests that, in these settings, the robust policy achieves
mortality reduction by accepting a higher false-positive burden relative
to ENGAGE. This is consistent with the benefit--harm trade-off \citep{Toumazis2021RiskBasedFramework} underlying lung cancer screening decisions: more frequent or earlier LDCT screening
can increase opportunities for detecting undiagnosed cancers, but it also
exposes more cancer-free individuals to abnormal LDCT findings and
therefore increases false positives.

The smallest ambiguity radius \(\epsilon=0.10\) shows a distinct pattern for male heavy smokers. Across the three perturbation intensities examined, the robust policy yields negative \(\Delta\mathrm{FP}\) and positive \(\Delta\mathrm{LCD}\), in contrast to the pattern observed at
larger ambiguity radii, where reductions in LCDs are accompanied by increases in FPs. This indicates that the LCD and FP pattern at \(\epsilon=0.10\) differs from the dominant pattern observed for male heavy smokers at moderate and large ambiguity radii.

To examine whether this small-radius pattern is sensitive to the random seed used in the perturbation and simulation procedures, we repeat the male-cohort evaluation under three independent random seeds in Appendix~\ref{app:sanity_check}.
 Table~\ref{tab:sanity_seeds} shows
that, at \(\epsilon=0.10\), the robust policy consistently yields negative \(\Delta\mathrm{FP}\) and positive \(\Delta\mathrm{LCD}\) across the tested random seeds and perturbation intensities. This matches
the direction observed in the main analysis and suggests that the small-radius pattern is not specific to a single random seed. 
We therefore interpret the male-cohort result at
\(\epsilon=0.10\) as a localized pattern at the smallest tested ambiguity
radius, rather than as representative of the robust POMDP across the larger ambiguity radii examined in this study.
A full mechanistic explanation of why this divergence appears for male heavy smokers at the smallest ambiguity radius is beyond the scope of this study.

A sex-specific difference also appears in the efficiency of the LCD and FP
trade-off. We measure this efficiency by the number of additional false
positives required to avert one lung cancer death,
\(\frac{\Delta\mathrm{FP}}{|\Delta\mathrm{LCD}|}\), 
computed only for cells in which \(\Delta\mathrm{LCD}<0\) and both
\(\Delta\mathrm{LCD}\) and \(\Delta\mathrm{FP}\) are statistically
significant. A smaller value of this ratio indicates that fewer
additional false positives are incurred for each averted lung cancer
death.

Under this metric, the robust policy is more efficient for female heavy
smokers than for male heavy smokers. Across the seven female cells
meeting these criteria, the robust policy incurs on average approximately
107 additional false positives per averted lung cancer death, with values
ranging from around 78 to 124. Across the eleven male cells meeting the same
criteria, the corresponding ratio averages approximately 326 and ranges
from around 93 to 620. This indicates that, among cells with statistically
significant reductions in LCDs and increases in FPs, the robust policy averts
lung cancer deaths with a lower additional false-positive burden for
female heavy smokers than for male heavy smokers. Details are provided in Appendix~\ref{app: Screening Efficiency Metric}.

\subsection{Impact on the Recommended Screening Schedule and Screen-Detected Cancers}\label{sec: sdc}

We further examine the screening behavior underlying these outcomes. Specifically, we evaluate how the robust policy changes the recommended use of LDCT and how many lung cancers are detected through screening. We use the same Monte Carlo evaluation framework as in Section~\ref{sec:lcd_fp}. We consider two additional policy-level metrics: the lifetime number of LDCT screens per person and the number of screen-detected lung cancers (SDCs) per $100{,}000$ individuals. SDCs are defined as transitions from an undiagnosed cancer state to a diagnosed cancer state under the screening action, and are further classified as early-stage or advanced-stage detections.

Table~\ref{tab:screens_sdc} reports the lifetime number of LDCT screens per person and the number of screen-detected cancers. The robust policy generally recommends more LDCT screening than nominal ENGAGE, although the magnitude of the increase differs by sex, perturbation intensity, and ambiguity radius. For female heavy smokers, the increase is small at $\alpha=0.5$, with additional screening below $0.02$ screens per person for $\epsilon \ge 0.5$. The increase becomes larger under stronger perturbations, reaching approximately $0.43$ to $0.52$ additional screens per person at $\alpha=3.0$ and $\epsilon \ge 0.5$. For male heavy smokers, the robust policy adds approximately $0.50$ to $0.75$ screens per person for most settings with $\epsilon \ge 0.5$, although the increase is smaller at $\alpha=3.0$ for the two largest ambiguity radii. 
We also find that the ENGAGE baseline recommends more lifetime screens for female heavy smokers than for male heavy smokers, with approximately \(28.4\) versus \(24.9\) screens per person at \(\alpha=0.5\).

\begin{table}[htbp]
\centering
\caption{Lifetime LDCT screens per person and SDCs per 100{,}000 individuals under ENGAGE and the Robust POMDP, across
ambiguity radii $\epsilon$ and perturbation intensities $\alpha$.
\label{tab:screens_sdc}}

\footnotesize
\setlength{\tabcolsep}{4pt}
\begin{tabular}{clrrrrr}
\hline
$\alpha$ & $\epsilon$ & Screens/person & $\Delta$Screens [95\% CI]
& SDCs/100k & $\Delta$SDC [95\% CI] & $\Delta$SDC e/a \\
\hline
\multicolumn{7}{l}{\textbf{Female heavy smoker}}\\
\hline
\multirow{6}{*}{$0.5$} & ENGAGE & 28.360 & --- & 31{,}721 & --- & --- \\
    & 0.10 & 28.368 & $+0.008$ [$-0.003,+0.019$] & 31{,}735 & $+14$ [$-9,+38$]    & $+12/+2$ \\
    & 0.50 & 28.371 & $+0.011$ [$+0.000,+0.022$] & 31{,}775 & $+54$ [$+31,+79$]   & $+54/+0$ \\
    & 0.90 & 28.372 & $+0.012$ [$+0.002,+0.023$] & 31{,}817 & $+96$ [$+71,+122$]  & $+99/-3$ \\
    & 1.70 & 28.375 & $+0.015$ [$+0.005,+0.026$] & 31{,}816 & $+95$ [$+71,+118$]  & $+99/-4$ \\
    & 1.90 & 28.372 & $+0.012$ [$+0.002,+0.023$] & 31{,}804 & $+83$ [$+58,+108$]  & $+85/-2$ \\
\hline
\multirow{6}{*}{$1.0$}& ENGAGE & 28.227 & --- & 31{,}671 & --- & --- \\
    & 0.10 & 28.385 & $+0.158$ [$+0.116,+0.200$] & 31{,}674 & $+3$ [$-21,+27$]    & $+1/+2$ \\
    & 0.50 & 28.391 & $+0.164$ [$+0.122,+0.206$] & 31{,}722 & $+51$ [$+26,+77$]   & $+52/-1$ \\
    & 0.90 & 28.394 & $+0.167$ [$+0.125,+0.209$] & 31{,}753 & $+82$ [$+56,+108$]  & $+82/+0$ \\
    & 1.70 & 28.394 & $+0.167$ [$+0.125,+0.209$] & 31{,}776 & $+105$ [$+78,+132$] & $+106/-1$ \\
    & 1.90 & 28.392 & $+0.165$ [$+0.123,+0.206$] & 31{,}772 & $+101$ [$+78,+125$] & $+104/-3$ \\
\hline
\multirow{6}{*}{$3.0$} & ENGAGE & 28.106 & --- & 31{,}013 & --- & --- \\
    & 0.10 & 28.267 & $+0.161$ [$+0.105,+0.218$] & 31{,}108 & $+95$ [$+65,+125$]   & $+93/+2$ \\
    & 0.50 & 28.531 & $+0.425$ [$+0.361,+0.489$] & 31{,}265 & $+252$ [$+209,+295$] & $+251/+1$ \\
    & 0.90 & 28.620 & $+0.514$ [$+0.451,+0.577$] & 31{,}312 & $+299$ [$+249,+349$] & $+302/-3$ \\
    & 1.70 & 28.630 & $+0.524$ [$+0.461,+0.589$] & 31{,}389 & $+376$ [$+315,+437$] & $+380/-4$ \\
    & 1.90 & 28.627 & $+0.521$ [$+0.458,+0.585$] & 31{,}367 & $+354$ [$+295,+413$] & $+359/-5$ \\
\hline
\multicolumn{7}{l}{\textbf{Male heavy smoker}}\\
\hline
\multirow{6}{*}{$0.5$} & ENGAGE & 24.904 & --- & 25{,}999 & --- & --- \\
    & 0.10 & 24.646 & $-0.258$ [$-0.308,-0.208$] & 26{,}005 & $+6$ [$-16,+27$]    & $+7/-1$ \\
    & 0.50 & 25.652 & $+0.748$ [$+0.698,+0.799$] & 26{,}065 & $+66$ [$+43,+87$]   & $+56/+10$ \\
    & 0.90 & 25.652 & $+0.748$ [$+0.698,+0.798$] & 26{,}066 & $+67$ [$+45,+87$]   & $+56/+11$ \\
    & 1.70 & 25.650 & $+0.746$ [$+0.695,+0.796$] & 26{,}077 & $+78$ [$+55,+99$]   & $+66/+12$ \\
    & 1.90 & 25.642 & $+0.738$ [$+0.688,+0.789$] & 26{,}080 & $+81$ [$+59,+103$]  & $+67/+14$ \\
\hline
\multirow{6}{*}{$1.0$} & ENGAGE & 25.016 & --- & 25{,}978 & --- & --- \\
    & 0.10 & 24.665 & $-0.351$ [$-0.405,-0.296$] & 25{,}965 & $-13$ [$-36,+9$]    & $-12/-1$ \\
    & 0.50 & 25.665 & $+0.649$ [$+0.595,+0.704$] & 26{,}046 & $+68$ [$+44,+91$]   & $+61/+7$ \\
    & 0.90 & 25.672 & $+0.656$ [$+0.602,+0.711$] & 26{,}063 & $+85$ [$+61,+108$]  & $+77/+8$ \\
    & 1.70 & 25.551 & $+0.535$ [$+0.478,+0.592$] & 26{,}037 & $+59$ [$+36,+82$]   & $+52/+7$ \\
    & 1.90 & 25.525 & $+0.509$ [$+0.452,+0.566$] & 26{,}073 & $+95$ [$+73,+118$]  & $+89/+6$ \\
\hline
\multirow{6}{*}{$3.0$} & ENGAGE & 25.073 & --- & 25{,}651 & --- & --- \\
    & 0.10 & 24.621 & $-0.452$ [$-0.521,-0.382$] & 25{,}701 & $+50$ [$+23,+77$]    & $+51/-1$ \\
    & 0.50 & 25.656 & $+0.583$ [$+0.508,+0.658$] & 25{,}862 & $+211$ [$+166,+257$] & $+207/+4$ \\
    & 0.90 & 25.793 & $+0.720$ [$+0.634,+0.806$] & 25{,}909 & $+258$ [$+208,+309$] & $+253/+5$ \\
    & 1.70 & 25.406 & $+0.333$ [$+0.259,+0.408$] & 25{,}946 & $+295$ [$+238,+353$] & $+295/+0$ \\
    & 1.90 & 25.398 & $+0.325$ [$+0.252,+0.399$] & 25{,}928 & $+277$ [$+222,+332$] & $+278/-1$ \\
\hline
\end{tabular}

\vspace{0.8em}

\begin{minipage}{\linewidth}
\footnotesize
Values are means over $M=300$ perturbation
realizations, each simulating $N=100{,}000$ individuals; brackets give 95\%
confidence intervals. $\Delta$Screens and $\Delta$SDC compare Robust vs.\
ENGAGE under identical perturbations. SDC counts and their differences are
reported as integers; $\Delta$SDC is split into early-stage
($s_2\!\to\!s_4$) and advanced-stage ($s_3\!\to\!s_5$) detections.
\end{minipage}

\end{table}

Table~\ref{tab:screens_agebin} decomposes the number of screens by age group at a representative setting  ($\alpha=0.5$ and $\epsilon=1.70$). For male heavy smokers, the additional screening under the robust policy is concentrated between ages $50$ and $59$, where the robust policy adds about $0.74$ screens per person. Differences between the two policies are negligible in the older age groups. For female heavy smokers, the schedule changes are small across all age groups, with a total increase of only $0.015$ screens per person. The small positive difference in the age $90$ to $100$ group lies close to the finite-horizon terminal age and is not interpreted as a clinically meaningful recommendation.

\begin{table}[htbp]
\centering
\caption{Age distribution of LDCT screens per person under ENGAGE and the
Robust POMDP at the representative setting $\alpha=0.5$, $\epsilon=1.70$,
for female and male heavy smokers. 
\label{tab:screens_agebin}}

\footnotesize
\begin{tabular}{lrrr}
\hline
Age group & ENGAGE & Robust & $\Delta$Screens [95\% CI] \\
\hline
\multicolumn{4}{l}{\textbf{Female heavy smoker}}\\
\hline
50--59  & 9.826  & 9.833  & $+0.007$ [$-0.002,+0.016$] \\
60--69  & 8.915  & 8.915  & $+0.000$ [$-0.001,+0.002$] \\
70--79  & 6.586  & 6.587  & $+0.001$ [$-0.002,+0.003$] \\
80--89  & 2.778  & 2.777  & $-0.001$ [$-0.002,+0.001$] \\
90--100 & 0.255  & 0.263  & $+0.008$ [$+0.007,+0.009$] \\
TOTAL   & 28.360 & 28.375 & $+0.015$ \\
\hline
\multicolumn{4}{l}{\textbf{Male heavy smoker}}\\
\hline
50--59  & 9.024  & 9.764  & $+0.740$ [$+0.690,+0.790$] \\
60--69  & 8.548  & 8.548  & $+0.000$ [$-0.002,+0.002$] \\
70--79  & 5.614  & 5.614  & $+0.000$ [$-0.002,+0.003$] \\
80--89  & 1.662  & 1.663  & $+0.001$ [$-0.001,+0.002$] \\
90--100 & 0.056  & 0.061  & $+0.005$ [$+0.004,+0.005$] \\
TOTAL   & 24.904 & 25.650 & $+0.746$ \\
\hline
\end{tabular}

\vspace{0.8em}

\begin{minipage}{\linewidth}
\footnotesize
Values are means over $M=300$
perturbation realizations; brackets give 95\% confidence intervals.
$\Delta$Screens compares Robust vs.\ ENGAGE under identical perturbations.
\end{minipage}

\end{table}

Table~\ref{tab:screens_sdc} also shows that, for ambiguity radii $\epsilon \ge 0.5$, the robust policy detects more lung cancers than ENGAGE in both cohorts. The increase ranges from approximately $51$ to $376$ additional screen-detected cancers per $100{,}000$ female heavy smokers and from approximately $59$ to $295$ additional screen-detected cancers per $100{,}000$ male heavy smokers. These additional detections are predominantly early-stage cancers, whereas the changes in advanced-stage screen-detected cancers are much smaller. Thus, the QALY improvements observed in Sections~\ref{sec:impact_radius} and~\ref{sec:impact_alpha} are consistent with more active screening and increased early-stage cancer detection.
The male results at $\epsilon=0.10$ should be interpreted separately from the larger-radius cases. At this radius, the robust policy recommends fewer screens than ENGAGE across all three perturbation intensities, and the change in screen-detected cancers is smaller than in the larger-radius cases and does not remain consistently positive across perturbation intensities. Therefore, the mechanism described above is most evident for ambiguity radii $\epsilon \ge 0.5$. 

The screening-pattern results provide a policy-level interpretation of the clinical outcome results reported in Section~\ref{sec:lcd_fp}. Compared with nominal ENGAGE, the robust policy generally recommends more LDCT screens and detects more early-stage cancers, which is consistent with the observed QALY improvements and reductions in lung cancer deaths. Additional screening also exposes more cancer-free individuals to LDCT and contributes to the increase in false positives. These results indicate that, for the main settings with $\epsilon \ge 0.5$, the robust policy's out-of-sample performance improvements are primarily associated with more active screening and earlier cancer detection.


\section{Conclusion}

In this paper, we develop a robust partially observable Markov decision
process (POMDP) framework for personalized lung cancer screening under
uncertainty in the cancer-state transition probabilities. 
To account for this uncertainty, we construct $\ell_1$-norm ambiguity sets around the nominal cancer-state transition kernel and seek the optimal screening policy that hedges against the worst-case kernel within these sets. Building on the piecewise-linear and convex (PWLC) structure of
the resulting robust value function, we adapt a point-based value iteration method \citep{osogami15} to compute robust screening policies.

We evaluated the proposed robust policies through out-of-sample
simulations in which selected cancer-progression parameters were perturbed using mean-corrected log-normal multipliers. The evaluation compared policies derived from the robust POMDP model with the nominal ENGAGE policy \citep{Toumazis2021RiskBasedFramework} for representative female and male heavy-smoker cohorts at age 50. The results show that robust POMDP policies generally outperform the
nominal ENGAGE policy in mean out-of-sample QALYs when the cancer-state transition model is perturbed.

We further evaluated clinical outcome measures that complement the QALY-based analysis, including lung cancer deaths (LCDs) and false positives (FPs) per 100{,}000 individuals. The robust policy reduces LCDs relative to the nominal ENGAGE policy in all evaluated settings for the female cohort and in most evaluated settings for the male cohort. In the settings where LCDs are reduced relative to the nominal ENGAGE
policy, the robust policy also produces more FPs, indicating that its mortality benefit is obtained at the cost of additional false-positive findings.
Among settings with statistically significant LCD reductions and FP increases, the robust policy achieves a lower additional FP burden per averted lung
cancer death for female heavy smokers than for male heavy smokers.

We also examined the screening schedules and screen-detected cancers generated by the robust policy to clarify the mechanism underlying these clinical outcome patterns. The robust policy generally recommends more lifetime LDCT screens than the nominal ENGAGE policy and, for ambiguity radii $\epsilon \ge 0.5$, detects more lung cancers, with most of the additional detections occurring at an early stage. These results suggest that the QALY gains and LCD reductions are primarily associated with more active screening and earlier cancer detection, whereas the additional screening contributes to the increase in false positives.

These findings suggest that robust POMDPs provide a useful extension of nominal POMDP-based lung cancer screening models by incorporating robustness against uncertainties in the transition kernel. Future work should focus on data-driven procedures for calibrating
transition-probability ambiguity sets and on evaluating the proposed framework in larger and more diverse screening cohorts.

\bibliographystyle{apalike}
\spacingset{1}
\bibliography{IISE-Trans}

@article{Thandra2021LungCancer,
  author    = {Thandra, K. C. and Barsouk, A. and Saginala, K. and Aluru, J. S. and Barsouk, A.},
  title     = {Epidemiology of lung cancer},
  journal   = {Contemporary Oncology (Poznan, Poland)},
  year      = {2021},
  volume    = {25},
  number    = {1},
  pages     = {45--52},
}

@article{Thakur2020LungCancer,
  author    = {Thakur, S. K. and Singh, D. P. and Choudhary, J.},
  title     = {Lung cancer identification: A review on detection and classification},
  journal   = {Cancer Metastasis Reviews},
  year      = {2020},
  volume    = {39},
  number    = {3},
  pages     = {989--998},
}

@article{NLST2011ReducedMortality,
  author  = {{National Lung Screening Trial Research Team}},
  title   = {Reduced Lung-Cancer Mortality with Low-Dose Computed Tomographic Screening},
  journal = {New England Journal of Medicine},
  volume  = {365},
  number  = {5},
  pages   = {395--409},
  year    = {2011},
  doi     = {10.1056/NEJMoa1102873}
}

@article{Steimle2021MultiModelMDP,
  author    = {Steimle, L. N. and Kaufman, D. L. and Denton, B. T.},
  title     = {Multi-model Markov decision processes},
  journal   = {IISE Transactions},
  year      = {2021},
  volume    = {53},
  number    = {10},
  pages     = {1124--1139},
}

@article{Toumazis2021RiskBasedFramework,
  author    = {Toumazis, I. and Alagoz, O. and Leung, A. and Plevritis, S. K.},
  title     = {A risk-based framework for assessing real-time lung cancer screening eligibility that incorporates life expectancy and past screening findings},
  journal   = {Cancer},
  year      = {2021},
  volume    = {127},
  number    = {23},
  pages     = {4432--4446},
  doi       = {10.1002/cncr.33846},
  note      = {Epub 2021 Aug 12. PMID: 34383299; PMCID: PMC8578300},
}

@InProceedings{osogami15,
  title = 	 {Robust partially observable Markov decision process},
  author = 	 {Osogami, Takayuki},
  booktitle = 	 {Proceedings of the 32nd International Conference on Machine Learning},
  pages = 	 {106--115},
  year = 	 {2015},
  editor = 	 {Bach, Francis and Blei, David},
  volume = 	 {37},
  series = 	 {Proceedings of Machine Learning Research},
  address = 	 {Lille, France},
  month = 	 {07--09 Jul},
  publisher =    {PMLR},
  url = 	 {https://proceedings.mlr.press/v37/osogami15.html}
}

@book{pomdp_book, 
place={Cambridge}, title={Partially Observed Markov Decision Processes: From Filtering to Controlled Sensing}, publisher={Cambridge University Press}, author={Krishnamurthy, Vikram}, year={2016}}

@article{Bach2003,
  author    = {Peter B. Bach and Mitchell W. Kattan and Mary D. Thornquist and Mark G. Kris and Richard C. Tate and Mark J. Barnett and Linda J. Hsieh and Colin B. Begg},
  title     = {Variations in lung cancer risk among smokers},
  journal   = {Journal of the National Cancer Institute},
  volume    = {95},
  number    = {6},
  pages     = {470--478},
  year      = {2003},
  doi       = {10.1093/jnci/95.6.470},
  publisher = {Oxford University Press}
}

@misc{acr2014lungrads,
  author       = {{American College of Radiology}},
  title        = {{Lung CT Screening Reporting and Data System (Lung-RADS)}},
  year         = {2014},
  howpublished = {\url{http://www.acr.org/Quality-Safety/Resources/LungRADS}},
  note         = {[Internet]}
}

@article{walter2015symptoms,
  author  = {Walter, F. M. and Rubin, G. and Bankhead, C. and Morris, H. C. and Hall, N. and Mills, K. and others},
  title   = {Symptoms and other factors associated with time to diagnosis and stage of lung cancer: a prospective cohort study},
  journal = {Br J Cancer},
  year    = {2015},
  volume  = {112},
  pages   = {S6--S13},
  doi     = {10.1038/bjc.2015.30},
  url     = {http://dx.doi.org/10.1038/bjc.2015.30}
}

@article{han2017compliance,
  author  = {Han, S. S. and Erdogan, S. A. and Toumazis, I. and Leung, A. and Plevritis, S. K.},
  title   = {Evaluating the impact of varied compliance to lung cancer screening recommendations using a microsimulation model},
  journal = {Cancer Causes \& Control},
  year    = {2017},
  volume  = {28},
  pages   = {947--958},
  doi     = {10.1007/s10552-017-0907-x}
}

@article{jeon2012actual,
   title={Actual and counterfactual smoking prevalence rates in the {U.S.} population via microsimulation},
   author={Jeon, Jihyoun and Meza, Rafael and Krapcho, Martin and Clarke, Lauren D and Byrne, Jeff and Levy, David T},
   journal={Risk Analysis},
   volume={32},
   number={S1},
   pages={S51--S68},
   year={2012},
   publisher={Wiley},
   doi={10.1111/j.1539-6924.2011.01775.x}
 }

@article{holford2014patterns,
   title={Patterns of birth cohort--specific smoking histories, 1965--2009},
   author={Holford, Theodore R and Levy, David T and McKay, Lisa A and Clarke, Lauren and Racine, Ben and Meza, Rafael and Land, Stephanie and Jeon, Jihyoun and Feuer, Eric J},
   journal={American Journal of Preventive Medicine},
   volume={46},
   number={2},
   pages={e31--e37},
   year={2014},
   publisher={Elsevier},
   doi={10.1016/j.amepre.2013.10.022}
 }

@article{rosenberg2012cohort,
   title={Cohort life tables by smoking status, removing lung cancer as a cause of death},
   author={Rosenberg, Mark A and Feuer, Eric J and Yu, Binbing and Sun, Jiafeng and Henley, Jane and Shanks, Thomas and Anderson, Christy M and McMahon, Pamela M and Thun, Michael and Burns, David M},
   journal={Risk Analysis},
   volume={32},
   number={S1},
   pages={S25--S38},
   year={2012},
   publisher={Wiley},
   doi={10.1111/j.1539-6924.2011.01759.x}
}

@article{iyengar,
  title={Robust dynamic programming},
  author={Iyengar, Garud N},
  journal={Mathematics of Operations Research},
  volume={30},
  number={2},
  pages={257--280},
  year={2005},
  publisher={INFORMS}
}

@article{nilim,
    title={Robut control of Markov decision processes with uncertain transition matrices.},
    author={Nilim, Arnab and El Ghaoui, Laurent},
    journal={ Operations Research},
    volume={53},
    pages={780--798},
    year={2005}
}

@incollection{Alagoz2014TutOR,
  author    = {Alagoz, Oguzhan},
  title     = {Optimizing Cancer Screening Using Partially Observable Markov Decision Processes},
  booktitle = {INFORMS TutORials in Operations Research},
  pages     = {75--89},
  year      = {2014},
  publisher = {INFORMS},
  doi       = {10.1287/educ.1110.0087}
}

@article{MaillartIvyRansomDiehl2008,
  author  = {Maillart, Lisa M. and Ivy, Julie Simmons and Ransom, Scott and Diehl, Kathleen},
  title   = {Assessing Dynamic Breast Cancer Screening Policies},
  journal = {Operations Research},
  volume  = {56},
  number  = {6},
  pages   = {1411--1427},
  year    = {2008},
  doi     = {10.1287/opre.1080.0614}
}

@article{AyerAlagozStoutBurnside2016,
  author  = {Ayer, Turgay and Alagoz, Oguzhan and Stout, Natasha K. and Burnside, Elizabeth S.},
  title   = {Heterogeneity in Women's Adherence and Its Role in Optimal Breast Cancer Screening Policies},
  journal = {Management Science},
  volume  = {62},
  number  = {5},
  pages   = {1339--1362},
  year    = {2016}
}

@techreport{SandikciCevikSchacht2018,
  author      = {Sand{\i}k{\c{c}}{\i}, Burhaneddin and Cevik, Mucahit and Schacht, David V.},
  title       = {Screening for Breast Cancer: The Role of Supplemental Tests and Breast Density Information},
  institution = {University of Chicago Booth School of Business},
  type        = {Working Paper},
  number      = {18-03},
  year        = {2018},
  note        = {SSRN Working Paper 3122581}
}

@article{ErenayAlagozSaid2014,
  author  = {Erenay, Fatih Safa and Alagoz, Oguzhan and Said, Adnan},
  title   = {Optimizing Colonoscopy Screening for Colorectal Cancer Prevention and Surveillance},
  journal = {Manufacturing \& Service Operations Management},
  volume  = {16},
  number  = {3},
  pages   = {381--400},
  year    = {2014},
  doi     = {10.1287/msom.2014.0484}
}

@article{KamalzadehAhujaHahslerBowen2021,
  author  = {Kamalzadeh, Hossein and Ahuja, Vishal and Hahsler, Michael and Bowen, Michael E.},
  title   = {An Analytics-Driven Approach for Optimal Individualized Diabetes Screening},
  journal = {Production and Operations Management},
  volume  = {30},
  number  = {9},
  pages   = {3161--3191},
  year    = {2021},
  doi     = {10.1111/poms.13422}
}

@article{WuSuen2022,
  author  = {Wu, Chou-Chun and Suen, Sze-chuan},
  title   = {Optimizing Diabetes Screening Frequencies for At-Risk Groups},
  journal = {Health Care Management Science},
  volume  = {25},
  pages   = {1--23},
  year    = {2022},
  doi     = {10.1007/s10729-021-09575-z}
}

@article{WuCaoSuenLin2024,
  author  = {Wu, Chou-Chun and Cao, Yiwen and Suen, Sze-chuan and Lin, Eugene},
  title   = {Examining Chronic Kidney Disease Screening Frequency Among Diabetics: A POMDP Approach},
  journal = {Health Care Management Science},
  volume  = {27},
  pages   = {391--414},
  year    = {2024},
  doi     = {10.1007/s10729-024-09677-4}
}

@article{SuenBrandeauGoldhaberFiebert2018,
  author  = {Suen, Sze-chuan and Brandeau, Margaret L. and Goldhaber-Fiebert, Jeremy D.},
  title   = {Optimal Timing of Drug Sensitivity Testing for Patients on First-Line Tuberculosis Treatment},
  journal = {Health Care Management Science},
  volume  = {21},
  pages   = {632--646},
  year    = {2018},
  doi     = {10.1007/s10729-017-9416-4}
}

@article{Petousis2019SequentialDecision,
  author  = {Petousis, Panayiotis and Winter, Audrey and Speier, William and Aberle, Denise R. and Hsu, William and Bui, Alex A. T.},
  title   = {Using Sequential Decision Making to Improve Lung Cancer Screening Performance},
  journal = {IEEE Access},
  year    = {2019},
  volume  = {7},
  pages   = {119403--119419},
  doi     = {10.1109/ACCESS.2019.2935763}
}

@phdthesis{Han2023MultipartPOMDP,
  author = {Han, Simon X.},
  title  = {A Multi-part Optimization Framework for POMDPs in Lung Cancer Screening},
  school = {University of California, Los Angeles},
  year   = {2023},
  url    = {https://escholarship.org/uc/item/6t43g1n2}
}

@inproceedings{Wang2024DeepRLEarlyDiagnosis,
  author    = {Wang, Yifan and Zhang, Qining and Ying, Lei and Zhou, Chuan},
  title     = {Deep Reinforcement Learning for Early Diagnosis of Lung Cancer},
  booktitle = {Proceedings of the AAAI Conference on Artificial Intelligence},
  year      = {2024}
}

@article{Toyoda2008LDCTSensSpec,
  author  = {Toyoda, Y. and Nakayama, T. and Kusunoki, Y. and Iso, H. and Suzuki, T.},
  title   = {Sensitivity and specificity of lung cancer screening using chest low-dose computed tomography},
  journal = {British Journal of Cancer},
  year    = {2008},
  volume  = {98},
  pages   = {1602--1607},
  doi     = {10.1038/sj.bjc.6604351}
}

@article{Hamilton2005ClinicalFeaturesLungCancer,
  author  = {Hamilton, W. and Peters, T. J. and Round, A. and Sharp, D.},
  title   = {What are the clinical features of lung cancer before the diagnosis is made? A population based case-control study},
  journal = {Thorax},
  year    = {2005},
  volume  = {60},
  pages   = {1059--1065}
}

@misc{seer2023,
  author       = {{Surveillance, Epidemiology, and End Results (SEER) Program}},
  title        = {{SEER*Stat Database}},
  organization = {National Cancer Institute, DCCPS, Surveillance Research Program},
  year         = {2023},
  howpublished = {\url{https://www.seer.cancer.gov}}
}

@article{deKoning2020nelson,
  title={Reduced lung-cancer mortality with volume {CT} screening in a randomized trial},
  author={de Koning, Harry J and van der Aalst, Carlijn M and de Jong, Pim A and others},
  journal={New England Journal of Medicine},
  volume={382}, number={6}, pages={503--513}, year={2020}
}

@article{ariasUSLifeTables2018,
  title={United States life tables, 2015},
  author={Arias, Elizabeth and Xu, Jiaquan},
  journal={National Vital Statistics Reports},
  volume={67},
  number={7},
  pages={1--64},
  year={2018}
}

@article{ITOH2007453,
title = {Partially observable Markov decision processes with imprecise parameters},
journal = {Artificial Intelligence},
volume = {171},
number = {8},
pages = {453-490},
year = {2007},
issn = {0004-3702},
doi = {https://doi.org/10.1016/j.artint.2007.03.004},
url = {https://www.sciencedirect.com/science/article/pii/S0004370207000471},
author = {Hideaki Itoh and Kiyohiko Nakamura}
}

@article{SAGHAFIAN20181,
title = {Ambiguous partially observable Markov decision processes: Structural results and applications},
journal = {Journal of Economic Theory},
volume = {178},
pages = {1-35},
year = {2018},
issn = {0022-0531},
doi = {https://doi.org/10.1016/j.jet.2018.08.006},
url = {https://www.sciencedirect.com/science/article/pii/S0022053118304770},
author = {Soroush Saghafian}
}

@techreport{RasouliSaghafian2018,
  author      = {Rasouli, Mohammad and Saghafian, Soroush},
  title       = {Robust Partially Observable Markov Decision Processes},
  institution = {Harvard Kennedy School},
  type        = {HKS Faculty Research Working Paper Series},
  number      = {RWP18-027},
  month       = sep,
  year        = {2018},
  url         = {https://www.hks.harvard.edu/publications/robust-partially-observable-markov-decision-processes}
}

@article{Boloori2020,
  author  = {Boloori, Alireza and Saghafian, Soroush and Chakkera, Harini A. and Cook, Curtiss B.},
  title   = {Data-Driven Management of Post-transplant Medications: An Ambiguous Partially Observable Markov Decision Process Approach},
  journal = {Manufacturing \& Service Operations Management},
  volume  = {22},
  number  = {5},
  pages   = {1066--1087},
  year    = {2020},
  doi     = {10.1287/msom.2019.0797}
}

@article{nakao,
author = {Nakao, Hideaki and Jiang, Ruiwei and Shen, Siqian},
title = {Distributionally Robust Partially Observable Markov Decision Process with Moment-Based Ambiguity},
journal = {SIAM Journal on Optimization},
volume = {31},
number = {1},
pages = {461-488},
year = {2021},
doi = {10.1137/19M1268410},

URL = { 
    
        https://doi.org/10.1137/19M1268410
    
    

},
eprint = { 
    
        https://doi.org/10.1137/19M1268410
    
    

}
}

@article{li2026distributionally,
  title={Distributionally Robust Partially Observable Markov Decision Process with Distance-Based Ambiguity Sets},
  author={Li, Tong and Xiang, Yisha},
  journal={IISE Transactions},
  volume={58},
  number={4},
  pages={396--411},
  year={2026},
  publisher={Taylor \& Francis}
}

@article{deKoning2014,
  author  = {de Koning, Harry J. and Meza, Rafael and Plevritis, Sylvia K. and ten Haaf, Kevin and Munshi, Vidit N. and Jeon, Jihyoun and Erdogan, S. Alper and Kong, Chung Yin and Han, Summer S. and van Rosmalen, Joost and Choi, Stephanie E. and Pinsky, Paul F. and Berrington de Gonzalez, Amy and Berg, Christine D. and Black, William C. and Tammem{\"a}gi, Martin C. and Hazelton, William D. and Feuer, Eric J. and McMahon, Pamela M.},
  title   = {Benefits and Harms of Computed Tomography Lung Cancer Screening Strategies: A Comparative Modeling Study for the {U.S.} Preventive Services Task Force},
  journal = {Annals of Internal Medicine},
  year    = {2014},
  volume  = {160},
  number  = {5},
  pages   = {311--320},
  doi     = {10.7326/M13-2316},
  pmid    = {24379002},
  pmcid   = {PMC4116741}
}

@article{Jonas2021,
  author  = {Jonas, Daniel E. and Reuland, Daniel S. and Reddy, Shyamali M. and Nagle, Megan and Clark, Sarah D. and Weber, Rachel P. and Enyioha, Chineme and Malo, Thandiwe L. and Brenner, Allison T. and Armstrong, Cecilia and Coker-Schwimmer, Meera and Middleton, Jennifer C. and Voisin, Christophe and Harris, Russell P.},
  title   = {Screening for Lung Cancer With Low-Dose Computed Tomography: Updated Evidence Report and Systematic Review for the {US} Preventive Services Task Force},
  journal = {JAMA},
  year    = {2021},
  volume  = {325},
  number  = {10},
  pages   = {971--987},
  doi     = {10.1001/jama.2021.0377}
}

@article{Meza2021,
  author  = {Meza, Rafael and Jeon, Jihyoun and Toumazis, Iakovos and Ten Haaf, Kevin and Cao, Peipei and Bastani, Meysam and Han, Summer S. and Blom, Elisabeth F. and Jonas, Daniel E. and Feuer, Eric J. and de Koning, Harry J. and Plevritis, Sylvia K. and Criss, Sean D.},
  title   = {Evaluation of the Benefits and Harms of Lung Cancer Screening With Low-Dose Computed Tomography: Modeling Study for the {US} Preventive Services Task Force},
  journal = {JAMA},
  year    = {2021},
  volume  = {325},
  number  = {10},
  pages   = {988--997},
  doi     = {10.1001/jama.2021.1077}
}

@misc{SEERLungCancerStats,
  author       = {{National Cancer Institute}},
  title        = {{SEER Cancer Stat Facts: Lung and Bronchus Cancer}},
  howpublished = {Surveillance, Epidemiology, and End Results Program},
  year         = {2026},
  url          = {https://seer.cancer.gov/statfacts/html/lungb.html},
  note         = {Accessed May 26, 2026}
}

@article{Church2013InitialLDCT,
  author  = {Church, Thomas R. and Black, William C. and Aberle, Denise R. and Berg, Christine D. and Clingan, Kimberly L. and Duan, Fenghai and Fagerstrom, Rachel M. and Gareen, Ilana F. and Gierada, David S. and Jones, Gary C. and Mahon, Ila and Marcus, Pamela M. and Sicks, JoRean D. and Jain, Anil and Baum, Steven},
  title   = {Results of Initial Low-Dose Computed Tomographic Screening for Lung Cancer},
  journal = {New England Journal of Medicine},
  year    = {2013},
  volume  = {368},
  number  = {21},
  pages   = {1980--1991},
  doi     = {10.1056/NEJMoa1209120},
  pmid    = {23697514},
  pmcid   = {PMC3762603}
}

@article{Aberle2013IncidenceScreenings,
  author  = {Aberle, Denise R. and DeMello, Stacie and Berg, Christine D. and Black, William C. and Brewer, Brenda and Church, Thomas R. and Clingan, Kimberly L. and Duan, Fenghai and Fagerstrom, Rachel M. and Gareen, Ilana F. and Gatsonis, Constantine A. and Gierada, David S. and Jain, Anil and Jones, Gary C. and Mahon, Ila and Marcus, Pamela M. and Rathmell, James M. and Sicks, JoRean D.},
  title   = {Results of the Two Incidence Screenings in the National Lung Screening Trial},
  journal = {New England Journal of Medicine},
  year    = {2013},
  volume  = {369},
  number  = {10},
  pages   = {920--931},
  doi     = {10.1056/NEJMoa1208962},
  pmid    = {24004119},
  pmcid   = {PMC4307922}
}

@article{USPSTF2021LungCancerScreening,
  author  = {{U.S. Preventive Services Task Force}},
  title   = {Screening for Lung Cancer: {US} Preventive Services Task Force Recommendation Statement},
  journal = {JAMA},
  year    = {2021},
  volume  = {325},
  number  = {10},
  pages   = {962--970},
  doi     = {10.1001/jama.2021.1117}
}

@article{Wolf2024ACSLungCancerScreening,
  author  = {Wolf, Andrew M. D. and Oeffinger, Kevin C. and Shih, Ya-Chen Tina and Walter, Louise C. and Church, Timothy R. and Fontham, Elizabeth T. H. and Elkin, Elena B. and Etzioni, Ruth D. and Guerra, Carmen E. and Perkins, Rebecca B. and Kondo, Karli K. and Kratzer, Travis B. and Manassaram-Baptiste, Deana and Dahut, William L. and Smith, Robert A.},
  title   = {Screening for Lung Cancer: 2023 Guideline Update from the American Cancer Society},
  journal = {CA: A Cancer Journal for Clinicians},
  year    = {2024},
  volume  = {74},
  number  = {1},
  pages   = {50--81},
  doi     = {10.3322/caac.21811}
}

@article{Marshall2001DecisionAnalysis,
  author  = {Marshall, Deborah and Simpson, Kit N. and Earle, Craig C. and Chu, Chia-Wen},
  title   = {Economic Decision Analysis Model of Screening for Lung Cancer},
  journal = {European Journal of Cancer},
  year    = {2001},
  volume  = {37},
  number  = {14},
  pages   = {1759--1767},
  doi     = {10.1016/S0959-8049(01)00205-2},
  pmid    = {11549429}
}

@article{Manser2005LDCTCostEffectiveness,
  author  = {Manser, Renata and Dalton, Anna and Carter, Rob and Byrnes, Graham and Elwood, Mark and Campbell, David A.},
  title   = {Cost-Effectiveness Analysis of Screening for Lung Cancer with Low Dose Spiral {CT} (Computed Tomography) in the Australian Setting},
  journal = {Lung Cancer},
  year    = {2005},
  volume  = {48},
  number  = {2},
  pages   = {171--185},
  doi     = {10.1016/j.lungcan.2004.11.001},
  note    = {Epub 2005 Jan 4},
  pmid    = {15829317}
}

@article{bray2024global,
  title={Global cancer statistics 2022: {GLOBOCAN} estimates of incidence and mortality worldwide for 36 cancers in 185 countries},
  author={Bray, Freddie and Laversanne, Mathieu and Sung, Hyuna and Ferlay, Jacques and Siegel, Rebecca L. and Soerjomataram, Isabelle and Jemal, Ahmedin},
  journal={CA: A Cancer Journal for Clinicians},
  volume={74},
  number={3},
  pages={229--263},
  year={2024},
  publisher={Wiley Online Library}
}

@ARTICLE{Fan2023LungCancerMortality,
    
AUTHOR={Fan, Yaguang  and Jiang, Yong  and Gong, Lei  and Wang, Ying  and Su, Zheng  and Li, Xuebing  and Wu, Heng  and Pan, Hongli  and Wang, Jing  and Meng, Zhaowei  and Zhou, Qinghua  and Qiao, Youlin },
           
TITLE={Epidemiological and demographic drivers of lung cancer mortality from 1990 to 2019: results from the global burden of disease study 2019},
          
JOURNAL={Frontiers in Public Health},
          
VOLUME={Volume 11 - 2023},
  
YEAR={2023},
  
URL={https://www.frontiersin.org/journals/public-health/articles/10.3389/fpubh.2023.1054200},
  
DOI={10.3389/fpubh.2023.1054200},
  
ISSN={2296-2565}
  
}

@article{Li2025GlobalLungCancerTrends,
  author  = {Li, Zhen and Yu, Cheng and Hao, Jiaxin and Luo, Nan and Peng, Hao and Zhang, Jie and Pu, Qiang and Liu, Lunxu},
  title   = {Global Trends of Early, Middle, and Late-Onset Lung Cancer From 1990 to 2021: Results From the Global Burden of Disease Study 2021},
  journal = {Cancer Medicine},
  year    = {2025},
  volume  = {14},
  number  = {3},
  pages   = {e70639},
  doi     = {10.1002/cam4.70639},
  pmid    = {39918236},
  pmcid   = {PMC11803626}
}


\newpage
\section*{Appendix}
\setcounter{subsection}{0}
\renewcommand{\thesubsection}{A.\arabic{subsection}}
\setcounter{equation}{0}
\renewcommand{\theequation}{A.\arabic{equation}}
\renewcommand{\thelemma}{A.\arabic{lemma}}
\renewcommand{\thecorollary}{A.\arabic{corollary}}
\renewcommand{\baselinestretch}{1}

\subsection{Supplementary Tables for the Numerical Experiments} \label{app:supplementary-tables} \subsubsection{Summary of model inputs} \label{app:model-inputs} Table~\ref{tab:model_inputs} summarizes the main model inputs used in the numerical experiments, including the discount factor, planning horizon, transition models, observation probabilities, ambiguity-set radii, perturbation intensities, and reward functions. These inputs are based on the ENGAGE model, published clinical and epidemiologic sources, and the experimental design specified in the main text. 

\begin{table}[h]
\centering
\caption{Summary of model inputs used in the numerical experiments.
\label{tab:model_inputs}}

{
\renewcommand{\arraystretch}{1.15}
\setlength{\tabcolsep}{6pt}
\begin{tabular}{@{}p{0.27\linewidth} p{0.42\linewidth} p{0.25\linewidth}@{}}
\toprule
\textbf{Description} & \textbf{Parameters} & \textbf{Source}\\
\midrule
Discount factor & $\beta = 1/1.03 \approx 0.97$ & \cite{Toumazis2021RiskBasedFramework} \\
Decision epoch length & 1 year & \cite{Toumazis2021RiskBasedFramework}\\
Start age / end age & 50 / 100 & \cite{Toumazis2021RiskBasedFramework} \\
Initial belief & $b_0$, see Appendix \ref{appendix: initial belief state} & Bach + SEER \\
Cancer-state transitions & $\widehat P_t(s'\mid s,m,a)$, see Appendix \ref{app:cancer-state-P} & Bach + ENGAGE \citep{Toumazis2021RiskBasedFramework} \\
Smoking-status transitions & $\widehat Q_t(m'\mid m)$, see Appendix \ref{app: smoking P} & CISNET + SHG \citep{jeon2012actual} \\
Observation probabilities & $\widehat Z_t(o\mid s,m,a)$ with $\mathrm{Se}^{(m)}_{\text{LDCT}}$, $\mathrm{Sp}^{(m)}_{\text{LDCT}}$, $\mathrm{Se}_{\mathrm{hemo}}$ & \cite{Toyoda2008LDCTSensSpec}, \cite{Hamilton2005ClinicalFeaturesLungCancer} \\
Ambiguity set radius & $\epsilon \in \{0.1, 0.2, \dots, 1.9\}$ & Assumed \\
Perturbation intensity & $\alpha \in \{0.05, 0.1, 0.2, 0.5, 1, 2, 3\}$ & Assumed \\
Reward functions & $R_t(\cdot)$, $G_t(\cdot)$, see Appendix \ref{app: reward} & \cite{Toumazis2021RiskBasedFramework} \\
\bottomrule
\end{tabular}
}

\end{table}

\subsection{Parameterization of the Nominal Clinical Model}\label{app:nominal_model_details}
\subsubsection{Cancer state transition probabilities $\widehat{P}_{t}(s_{t+1}|s_{t}, m_{t}, a_{t})$}\label{app:cancer-state-P}
Following \cite{Toumazis2021RiskBasedFramework}, we assume that
cancer-state transitions occur at the end of each annual decision epoch. The nominal cancer-state transition probability is denoted by \(\widehat P_t(s'\mid s,m,a)\). Because our formulation distinguishes the event structure under waiting and LDCT screening, we write the
nominal transition matrix under waiting as \(\widehat P_t^{(a_2)}\) and the nominal transition matrix under screening as \(\widehat P_t^{(a_1)}\).

When the action \(a_t=a_2\) (wait) is selected, the individual does not
undergo LDCT screening during the current decision epoch. The transition
matrix follows the no-screen cancer-state transition structure in
\cite{Toumazis2021RiskBasedFramework}. Transitions from the cancer-free
state \(s_1\) are determined by the Bach lung cancer incidence model and
the Bach other-cause mortality model \citep{Bach2003}. Specifically, \(\widehat p_{12}^W(t)\) and \(\widehat p_{13}^W(t)\) denote the
one-year probabilities of developing undiagnosed early-stage and advanced-stage lung cancer, respectively, and
\(\widehat p_{16}^W(t)\) denotes the one-year probability of death from causes other than lung cancer.

Once an individual is in an undiagnosed cancer state, subsequent preclinical progression and lung-cancer mortality follow the
natural-history model used in
\cite{Toumazis2021RiskBasedFramework}. The probabilities
\(\widehat p_{23}^W(t)\), \(\widehat p_{26}^W(t)\), and
\(\widehat p_{36}^W(t)\)  are estimated from the simulated one-year event frequencies reported in Supplementary Appendix 3 and Supplementary Table 1 of \cite{Toumazis2021RiskBasedFramework}.
Therefore, the cancer-state transition probability at action $a_2$ is given by
\begin{equation}
\widehat P_t^{(a_2)}=
\begin{bmatrix}
\widehat p_{11}^W(t) & \widehat p_{12}^W(t) & \widehat p_{13}^W(t) & 0 & 0 & \widehat p_{16}^W(t)\\
0 & \widehat p_{22}^W(t) & \widehat p_{23}^W(t) & 0 & 0 & \widehat p_{26}^W(t)\\
0 & 0 & \widehat p_{33}^W(t) & 0 & 0 & \widehat p_{36}^W(t)\\
0 & 0 & 0 & 1 & 0 & 0\\
0 & 0 & 0 & 0 & 1 & 0\\
0 & 0 & 0 & 0 & 0 & 1
\end{bmatrix}.
\end{equation}
Here, \(\widehat p_{12}^W(t)\) and \(\widehat p_{13}^W(t)\) are the one-year probabilities of developing undiagnosed early- and advanced-stage lung cancer from the cancer-free state, respectively; \(\widehat p_{23}^W(t)\) is the one-year probability of progression from undiagnosed early-stage to undiagnosed advanced-stage lung cancer; and \(\widehat p_{16}^W(t)\), \(\widehat p_{26}^W(t)\), and \(\widehat p_{36}^W(t)\) are the corresponding one-year probabilities of transition to death. The diagonal entries are determined by row normalization, namely
\[
\widehat p_{11}^W(t)=1-\widehat p_{12}^W(t)-\widehat p_{13}^W(t)-\widehat p_{16}^W(t),\qquad
\widehat p_{22}^W(t)=1-\widehat p_{23}^W(t)-\widehat p_{26}^W(t),\qquad
\widehat p_{33}^W(t)=1-\widehat p_{36}^W(t).
\]
Moreover, the diagnosed states \(s_4\) and \(s_5\), as well as the death state \(s_6\), are absorbing. Once an individual enters one of these states, no further screening decisions are made.

When the action \(a_t=a_1\) (screen) is selected, the individual undergoes LDCT screening during the current epoch. In this case, the underlying disease still follows the same natural-history dynamics as above, but screening may reveal an existing undiagnosed cancer and move the individual into the corresponding diagnosed state. Because LDCT sensitivity is less than one, diagnosis does not occur with certainty even when the true state is cancerous. In particular, for an individual in \(s_2\), an abnormal screening result occurs with probability equal to the LDCT sensitivity, and under the assumption of perfect diagnostic work-up, such a detected case is transferred to the diagnosed early-stage state \(s_4\). Likewise, for an individual in \(s_3\), an abnormal screening result is confirmed as advanced-stage lung cancer and the individual transitions to \(s_5\). If the screening result is false negative, the individual remains in the undiagnosed process and continues to evolve according to the natural-history probabilities. By contrast, if the individual is in the cancer-free state \(s_1\), an abnormal LDCT result may occur due to imperfect specificity, but because the follow-up diagnostic work-up is assumed perfect, this corresponds only to a false-positive screening episode and does not move the individual into \(s_4\) or \(s_5\). Therefore, \(\widehat P_t^{(a_1)}(s_4\mid s_1)=\widehat P_t^{(a_1)}(s_5\mid s_1)=0\).

Under this construction, the cancer-state transition probability under screening is given by
\[
\widehat P_t^{(a_1)}=
\begin{bmatrix}
\widehat p_{11}^S(t) & \widehat p_{12}^S(t) & \widehat p_{13}^S(t) & 0 & 0 & \widehat p_{16}^S(t)\\
0 & \widehat p_{22}^S(t) & \widehat p_{23}^S(t) & \widehat p_{24}^S(t) & 0 & \widehat p_{26}^S(t)\\
0 & 0 & \widehat p_{33}^S(t) & 0 & \widehat p_{35}^S(t) & \widehat p_{36}^S(t)\\
0 & 0 & 0 & 1 & 0 & 0\\
0 & 0 & 0 & 0 & 1 & 0\\
0 & 0 & 0 & 0 & 0 & 1
\end{bmatrix},
\]
where the superscript \(S\) denotes the screen action. The new off-diagonal terms \(\widehat p_{24}^S(t)\) and \(\widehat p_{35}^S(t)\) represent diagnosis-induced transitions generated by screening. In accordance with the observation model and the perfect-work-up assumption, these probabilities are determined by the smoking-status-specific LDCT sensitivity:
\[
\widehat p_{24}^S(t)=\mathrm{Se}_{\mathrm{LDCT}}^{(m_t)},\qquad
\widehat p_{35}^S(t)=\mathrm{Se}_{\mathrm{LDCT}}^{(m_t)}.
\]
The remaining probabilities in rows two and three correspond to cases that are not detected by screening and therefore continue to evolve according to the same natural-history dynamics as under waiting. A convenient implementation is to redistribute the remaining probability mass proportionally to the wait-action transition probabilities, namely
\[
\widehat p_{22}^S(t)=\bigl(1-\widehat p_{24}^S(t)\bigr)\widehat p_{22}^W(t),\qquad
\widehat p_{23}^S(t)=\bigl(1-\widehat p_{24}^S(t)\bigr)\widehat p_{23}^W(t),\qquad
\widehat p_{26}^S(t)=\bigl(1-\widehat p_{24}^S(t)\bigr)\widehat p_{26}^W(t),
\]
and
\[
\widehat p_{33}^S(t)=\bigl(1-\widehat p_{35}^S(t)\bigr)p_{33}^W(t),\qquad
\widehat p_{36}^S(t)=\bigl(1-\widehat p_{35}^S(t)\bigr)\widehat p_{36}^W(t).
\]
For the cancer-free state \(s_1\), screening does not create diagnosed cancer states, so the first row remains the same as under waiting:
\[
\widehat p_{11}^S(t)=p_{11}^W(t),\qquad
\widehat p_{12}^S(t)=p_{12}^W(t),\qquad
\widehat p_{13}^S(t)=p_{13}^W(t),\qquad
\widehat p_{16}^S(t)=p_{16}^W(t).
\]

In summary, the wait-action transition matrix represents the natural evolution of the disease in the absence of screening, whereas the screen-action transition matrix augments the same natural-history dynamics with diagnosis-induced transitions from \(s_2\) to \(s_4\) and from \(s_3\) to \(s_5\). The resulting transition probabilities are time-inhomogeneous, since they depend on age through the Bach model and the age-specific natural-history estimates adopted from \cite{Toumazis2021RiskBasedFramework}, and they also depend on smoking status through the LDCT sensitivity used under screening.

\subsubsection{Smoking state transition probabilities $\widehat Q_{t}(m_{t+1}|m_{t})$}\label{app: smoking P}
We model age- and sex-specific transitions among four smoking states: \textit{former}, \textit{light} ($<$10 cigarettes/day), \textit{moderate} (10--20 cigarettes/day), and \textit{heavy} ($>$20 cigarettes/day). These transitions are estimated using a microsimulation model that generates individual smoking histories following the methodology of the Cancer Intervention and Surveillance Modeling Network (CISNET) and its Smoking History Generator (SHG)~\citep{jeon2012actual}.

Specifically, we independently simulate smoking histories for 100,000 U.S.\ males and 100,000 U.S.\ females from a representative birth cohort (1960). For each individual, the simulation proceeds in four steps consistent with the SHG framework~\citep{jeon2012actual}: (1) an age at smoking initiation is determined using annual initiation probabilities estimated from the National Health Interview Survey (NHIS) via age-period-cohort models~\citep{holford2014patterns}; (2) a smoking intensity category (light, moderate, or heavy) is assigned at initiation based on the NHIS cigarettes-per-day distribution, with intensity fixed from age 30 onward~\citep{jeon2012actual}; (3) an age at smoking cessation is determined using annual cessation probabilities derived from the NHIS~\citep{holford2014patterns}, where cessation is defined as sustained abstinence for at least two years with no subsequent relapse~\citep{jeon2012actual}; and (4) an age at death from causes other than lung cancer is determined using U.S.\ life tables adjusted for differential mortality by smoking status based on relative risk estimates from the Cancer Prevention Study II (CPS-II)~\citep{rosenberg2012cohort}.

Importantly, the cessation probabilities reported by \cite{holford2014patterns} represent the quitting behavior of the general population of current smokers at each age, including individuals who are relatively likely to quit. However, the population of interest in our screening model consists of ever-smokers who remain current smokers at age 50. This subgroup represents persistent smokers who did not quit during the younger years when cessation rates are highest, and therefore exhibits a substantially lower propensity to quit compared to the general smoking population.
To account for this selection effect, our simulation employs a two-phase cessation approach: Phase~1 (ages 15--49) uses the general population cessation rates from \cite{holford2014patterns}, which naturally filters out individuals who are more likely to quit, retaining only persistent smokers by age 50; Phase~2 (ages 50--100) applies substantially lower cessation rates that reflect the reduced quitting propensity of this selected subgroup, consistent with the ``hardening hypothesis'' in tobacco control research.

Smoking status is updated annually at the end of each decision epoch. At age 50, each surviving ever-smoker is classified into one of the four smoking states based on their current smoking status and intensity. For individuals with prior smoking status $m_{t-1} \in \{\text{Light, Moderate, Heavy}\}$, the probability of transitioning to status $m_t$ at age $t$ and sex $g$ is empirically derived from the simulated smoking trajectories:
\begin{equation}
\widehat Q(m_t \mid m_{t-1}, \text{age} = t, \text{sex} = g) = \frac{N_{m_{t-1} \to m_t}^{(t,g)}}{N_{m_{t-1}}^{(t,g)}},
\label{eq:smoking_transition}
\end{equation}
where $N_{m_{t-1} \to m_t}^{(t,g)}$ denotes the number of individuals of sex $g$ and age $t$ who transition from $m_{t-1}$ to $m_t$, and $N_{m_{t-1}}^{(t,g)}$ is the number of individuals alive and in state $m_{t-1}$ at age $t$.

For former smokers, we assume no relapse and model the former smoking state as absorbing~\citep{jeon2012actual}:
\begin{equation}
\widehat Q(m_t \mid m_{t-1} = \text{Former}) = 
\begin{cases} 
1, & m_t = \text{Former}, \\ 
0, & m_t \in \{\text{Light, Moderate, Heavy}\}.
\end{cases}
\label{eq:former_absorbing}
\end{equation}

Since smoking intensity is fixed from age 30 onward in our simulation~\citep{jeon2012actual}, current smokers can only transition to the former state or remain in their current intensity category. Transitions between intensity categories (e.g., light to moderate) do not occur after age 50, which is consistent with the SHG assumption that smoking dose stabilizes by age 30.

This formulation captures time-varying smoking behavior and its downstream impact on lung cancer risk and LDCT screening effectiveness throughout the screening horizon.

\subsubsection{Observation probabilities $\widehat Z_{t}(o_{t}|s_{t}, m_{t}, a_{t})$}\label{app: observation P}
At decision epoch $t$, the observation probability is defined as
\(\widehat Z_t(o_t \mid s_t, m_t, a_t),\)
where $s_t$ denotes the underlying cancer-related state, $m_t$ denotes the smoking-status state, $a_t$ is the selected action, and $o_t$ is the realized observation. 

When $a_t=a_1$ (LDCT screening), the observation space is $o_t \in \{\text{abnormal}, \text{normal}\}$. Following \cite{Toumazis2021RiskBasedFramework}, the probability of observing an abnormal LDCT result is defined by the smoking-status-specific sensitivity and specificity of LDCT. Specifically,
\[
\widehat Z_t(o_t=\text{abnormal}\mid s_t,m_t,a_t=a_1)=
\begin{cases}
\mathrm{Se}^{(m_t)}_{\mathrm{LDCT}}, & \text{if } s_t \in \{s_2,s_3\},\\[4pt]
1-\mathrm{Sp}^{(m_t)}_{\mathrm{LDCT}}, & \text{if } s_t=s_1,
\end{cases}
\]
and
\[
\widehat Z_t(o_t=\text{normal}\mid s_t,m_t,a_t=a_1)
=
1-\widehat Z_t(o_t=\text{abnormal}\mid s_t,m_t,a_t=a_1).
\]
Here, $\mathrm{Se}^{(m_t)}_{\mathrm{LDCT}}$ and $\mathrm{Sp}^{(m_t)}_{\mathrm{LDCT}}$ denote the sensitivity and specificity of LDCT corresponding to smoking-status group $m_t$, respectively. These values are taken from Supplementary Table 3 of \cite{Toumazis2021RiskBasedFramework}.

When $a_t=a_2$ (wait), the observation space is $o_t \in \{\text{hemoptysis}, \text{no hemoptysis}\}$. In \cite{Toumazis2021RiskBasedFramework}, the symptom observation model additionally incorporated age- and sex-specific incidence rates of hemoptysis. To simplify the present model, we do not introduce age-specific symptom probabilities and instead define the observation probabilities solely using the sensitivity and specificity of hemoptysis reported in Supplementary Table 5 of \cite{Toumazis2021RiskBasedFramework}. Specifically,
\[
\widehat Z_t(o_t=\text{hemoptysis}\mid s_t,a_t=a_2)=
\begin{cases}
\mathrm{Se}_{\mathrm{hemo}}, & \text{if } s_t \in \{s_2,s_3\},\\[4pt]
1-\mathrm{Sp}_{\mathrm{hemo}}, & \text{if } s_t=s_1,
\end{cases}
\]
and
\[
\widehat Z_t(o_t=\text{no hemoptysis}\mid s_t,a_t=a_2)
=
1-\widehat Z_t(o_t=\text{hemoptysis}\mid s_t,a_t=a_2).
\]
Here, $\mathrm{Se}_{\mathrm{hemo}}$ and $\mathrm{Sp}_{\mathrm{hemo}}$ denote the sensitivity and specificity of hemoptysis as a lung cancer symptom, respectively. Under this simplified specification, the observation probabilities under the wait action depend only on the underlying cancer state and do not vary by age or sex.


\subsubsection{Initial belief state.}\label{appendix: initial belief state}
At the first decision epoch, corresponding to age 50, the patient's true cancer state is not directly observable unless lung cancer has already been diagnosed. Therefore, the initial belief state is defined over the partially observable cancer-state subset
\(\mathcal S^{PO}=\{s_1,s_2,s_3\}\), where \(s_1\) denotes the cancer-free state, \(s_2\) denotes undiagnosed early-stage lung cancer, and \(s_3\) denotes undiagnosed advanced-stage lung cancer.

We construct the initial belief state using the Bach lung cancer risk prediction model. Specifically, for an individual with demographic and smoking-history characteristics available at age 50, we evaluate the Bach model to obtain the predicted overall probability of lung cancer at age 50, denoted by
\(\mathrm{Risk}^{\mathrm{Bach}}_{50}\). This quantity is interpreted as the initial probability that the individual has undiagnosed lung cancer at the beginning of the screening horizon.

Because the Bach model provides an overall lung cancer risk rather than stage-specific probabilities, we allocate this risk to the undiagnosed early-stage and advanced-stage cancer states using SEER-based stage distribution weights. Let \(w_E\) and \(w_A\) denote the proportions of lung cancer cases assigned to early-stage and advanced-stage disease, respectively, at the corresponding age and sex group, with
\[
w_E+w_A=1.
\]
The initial belief state over \((s_1,s_2,s_3)\) is then given by
\[
b_{50}
=
\left[
1-\mathrm{Risk}^{\mathrm{Bach}}_{50},\;
w_E\mathrm{Risk}^{\mathrm{Bach}}_{50},\;
w_A\mathrm{Risk}^{\mathrm{Bach}}_{50}
\right].
\]
Thus, the probability assigned to the cancer-free state is the complement of the Bach-predicted overall lung cancer risk, while the probability assigned to the two undiagnosed cancer states is obtained by splitting the predicted risk into early- and advanced-stage components according to the SEER-based stage distribution.

\subsubsection{Reward}\label{app: reward}
We adopt the same reward specification as in
\cite{Toumazis2021RiskBasedFramework}. Specifically, the patient's
total expected QALYs are composed of two parts: the immediate rewards
accrued over annual decision epochs and a lump-sum terminal reward
assigned when the process enters an absorbing state or reaches the
terminal age. We next describe how each of these reward components is
formulated.

Let $R_t(s_t, m_t, a_t, o_t)$ denote the immediate QALYs reward at
epoch $t$, where $s_t$ is the underlying cancer-related state, $m_t$
is the smoking-status state, $a_t$ is the selected action, and $o_t$
is the realized observation.

If the selected action is to defer screening ($a_t = a_2$), then
following \cite{Toumazis2021RiskBasedFramework}, the immediate reward
is computed using a half-cycle correction. In particular, an
individual who survives the entire decision epoch accrues one full
year of quality-adjusted survival, whereas an individual who dies
during the epoch is assumed to accrue one-half year of
quality-adjusted survival on average. Therefore, for
$s_t \in \mathcal{S}^{PO}$, the immediate reward under no screening is
defined as
\begin{equation}
R_t(s_t, m_t, a_2, o_t)
=
R_t^{\text{No Screen}}(s_t, m_t)
=
u_{s_t}\left[P_{\text{alive}}(s_t) + 0.5\, P_{\text{death}}(s_t)\right],
\end{equation}
where $u_{s_t}$ denotes the health utility associated with the
individual's cancer-related state and smoking status, based on the
utility values reported in Supplementary Tables~6--8 of
\cite{Toumazis2021RiskBasedFramework}, $P_{\text{alive}}(s_t)$ denotes
the probability of surviving the current decision epoch conditional on
state $s_t$, and
\[
P_{\text{death}}(s_t) = 1 - P_{\text{alive}}(s_t)
\]
denotes the corresponding probability of death during that epoch.
Equivalently, if the transition probability to the death state $s_6$
is explicitly available, then
\[
P_{\text{death}}(s_t) = P_t(s_6 \mid s_t, m_t, a_t),
\qquad
P_{\text{alive}}(s_t) = 1 - P_t(s_6 \mid s_t, m_t, a_t).
\]

If LDCT screening is performed ($a_t = a_1$), the immediate reward is
defined using the same survival-based QALY contribution as above, but
reduced by the disutility associated with the realized screening
outcome \citep{Toumazis2021RiskBasedFramework}. Let $d_{o_t}$ denote
the observation-dependent disutility corresponding to the observed
LDCT result, with values determined according to Supplementary Table~9
of \cite{Toumazis2021RiskBasedFramework}. Then, for
$s_t \in \mathcal{S}^{PO}$, the immediate reward under screening is
given by
\begin{equation}
R_t(s_t, m_t, a_1, o_t)
=
R_t^{\text{Screen}}(s_t, m_t, o_t)
=
R_t^{\text{No Screen}}(s_t, m_t) - d_{o_t}.
\end{equation}
Thus, the screening reward consists of the same survival-based QALY
contribution as in the no-screening case, reduced by the disutility
associated with the realized screening observation.


We next describe the lump-sum terminal rewards. Let $G_t(s)$ denote
the lump-sum reward received upon entering an absorbing state
$s \in \{s_4, s_5, s_6\}$ at epoch $t$, or upon reaching the terminal
age $T$ while in state $s$. Following
\cite{Toumazis2021RiskBasedFramework}, $G_t(s)$ corresponds to the
patient's expected survival time conditional on the absorbing event,
and is constructed separately for the diagnosed lung cancer states,
the death state, and the cancer-free terminal state.

For diagnosed lung cancer states $s_4$ and $s_5$, the lump-sum reward
is defined using the age- and sex-specific expected survival functions
reported in Supplementary Appendix~4 and Supplementary Figure~3 of
\cite{Toumazis2021RiskBasedFramework}. For early-stage screen-detected
lung cancer, \cite{Toumazis2021RiskBasedFramework} estimated the
age-specific expected survival using the CISNET microsimulation model
developed at Stanford University \citep{han2017compliance}. The
resulting age-specific survival values were then fitted using
sex-specific cubic polynomial regressions, yielding
\begin{equation}
L^{E}(a, \text{sex}) =
\begin{cases}
0.000185\,a^3 - 0.041686\,a^2 + 2.803008\,a - 48.407829, & \text{if sex = male},\\[4pt]
0.000179\,a^3 - 0.038505\,a^2 + 2.327440\,a - 26.582712, & \text{if sex = female},
\end{cases}
\end{equation}
where $a$ denotes the age at diagnosis. For advanced-stage lung
cancer, \cite{Toumazis2021RiskBasedFramework} used SEER-based
age-specific survival estimates together with their assumed screening
benefit, and then fitted sex-specific linear regressions, yielding
\begin{equation}
L^{A}(a, \text{sex}) =
\begin{cases}
-0.030924\,a + 3.173304, & \text{if sex = male},\\[4pt]
-0.038780\,a + 3.956674, & \text{if sex = female}.
\end{cases}
\end{equation}

For the death state $s_6$, no further QALYs are accrued, since a
deceased individual has zero expected survival time. The lump-sum
reward upon entering $s_6$ is therefore zero.

Combining these cases, the lump-sum terminal reward at any nonterminal
epoch $t < T$ is
\begin{equation}
G_t(s) =
\begin{cases}
L^{E}(50 + t,\, \text{sex}), & s = s_4,\\[4pt]
L^{A}(50 + t,\, \text{sex}), & s = s_5,\\[4pt]
0, & s = s_6.
\end{cases}
\end{equation}

At the terminal age $T$ (corresponding to age 100), a boundary reward
is assigned to individuals who remain in the partially observable
subset. Following Supplementary Table~10 of
\cite{Toumazis2021RiskBasedFramework}, the cancer-free terminal reward
at age 100 is the residual expected survival time obtained from the
U.S.\ life tables \citep{ariasUSLifeTables2018}. Specifically,
\begin{equation}
G_T(s) =
\begin{cases}
2.1 \text{ years}, & s = s_1,\ \text{sex = male},\\[4pt]
2.3 \text{ years}, & s = s_1,\ \text{sex = female},\\[4pt]
0, & s \in \{s_2, s_3\}.
\end{cases}
\end{equation}
Individuals who remain in undiagnosed lung cancer states $s_2$ or
$s_3$ at age 100 receive no additional terminal reward, consistent
with the modeling assumption in
\cite{Toumazis2021RiskBasedFramework} that the screening process is
not extended beyond the study horizon.


\subsection{Bach Model for Cancer-Free-State Transitions}
\label{app:bach_transition}

This appendix describes how we use the Bach model~\citep{Bach2003}
to compute the time-varying transition probabilities out of the cancer-free
state. These probabilities determine the first row of the cancer-state
transition matrix at each decision epoch \(t\).

At each decision epoch \(t\), we compute two Bach one-year risks: 
(i) the one-year probability of incident lung cancer, denoted by
\(\lambda_t\), and 
(ii) the one-year probability of death in the absence of a lung cancer
diagnosis, denoted by \(\mu_t\). These risks are obtained from the
supplementary proportional-hazards equations of
\citet{Bach2003}, which model lung cancer incidence and
competing mortality as functions of smoking intensity, smoking duration,
years since quitting, age, asbestos exposure, and sex. For current
smokers, years since quitting is set to zero.

Specifically, we compute
\begin{equation}
\lambda_t \;=\; 1 - 0.99629^{\exp\bigl(\eta_{\mathrm{LC}}(t)\bigr)},
\qquad
\mu_t \;=\; 1 - 0.9917663^{\exp\bigl(\eta_{\mathrm{Death}}(t)\bigr)},
\label{eq:bach_lambda_mu}
\end{equation}
where \(\eta_{\mathrm{LC}}(t)\) and \(\eta_{\mathrm{Death}}(t)\) denote the
Bach one-year lung cancer incidence and competing-mortality linear
predictors, respectively, evaluated at the covariate profile for epoch
\(t\). We use the coefficient values reported in the supplementary
material of \citet{Bach2003}. Our implementation follows the
same Bach-model specification adopted in prior POMDP-based lung cancer
screening studies~\citep{Toumazis2021RiskBasedFramework}.

Using \(\lambda_t\) and \(\mu_t\) from~\eqref{eq:bach_lambda_mu}, we define
the transition probabilities out of the cancer-free state at epoch \(t\)
as
\begin{equation}
p_{12}(t)=\lambda_t,\qquad
p_{16}(t)=\mu_t,\qquad
p_{11}(t)=1-\lambda_t-\mu_t.
\label{eq:first_row}
\end{equation}
We set \(p_{13}(t)=0\) by construction, so that all incident cancers enter
the undiagnosed early-stage cancer state. Therefore, the first row of the
cancer-state transition matrix at epoch \(t\) is
\begin{equation}
P_t(s_1,\cdot)
=\bigl[\,1-\lambda_t-\mu_t,\ \lambda_t,\ 0,\ 0,\ 0,\ \mu_t\,\bigr].
\label{eq:first_row_vector}
\end{equation}
\subsection{Out-of-Sample Perturbation Design}
\label{app:perturbation_design}

This appendix describes the perturbation procedure used in the out-of-sample evaluation. 
For each Monte Carlo sample \(i=1,\ldots,M\), we generate one perturbed
out-of-sample environment by constructing a perturbed cancer-state
transition kernel \(\widehat P_t^{(i)}\). We perturb selected underlying parameters used to construct
\(\widehat P_t\), rather than perturbing every entry of the transition matrix directly. The selected
parameters are
\(
\theta \in \{\lambda_t, \mu_t, p_{23}, p_{26}, p_{36}\}.
\)
Here, \(\lambda_t\) denotes the annual lung cancer onset probability,
\(\mu_t\) denotes the competing-mortality probability used in the
cancer-state transition model, and \(p_{23}\), \(p_{26}\), and \(p_{36}\)
denote natural-history transition probabilities out of the preclinical
cancer states.

For each selected parameter \(\theta\) and each Monte Carlo sample \(i\), we define
\begin{equation}
  \theta^{(i)}
  =
  \rho_\theta^{(i)} \theta_{\mathrm{nom}},
  \qquad
  \rho_\theta^{(i)}
  =
  \exp\!\left(
  \alpha \sigma_\theta \xi_\theta^{(i)}
  -
  \frac{1}{2}\alpha^2\sigma_\theta^2
  \right),
  \qquad
  \xi_\theta^{(i)} \sim \mathcal{N}(0,1).
  \label{eq:log_norm_perturbation}
\end{equation}
In \eqref{eq:log_norm_perturbation}, \(\theta_{\mathrm{nom}}\) is the
corresponding nominal parameter value and \(\rho_\theta^{(i)}\) is the mean-corrected log-normal multiplier applied to parameter \(\theta\) in
Monte Carlo sample \(i\). The hyperparameter \(\sigma_\theta\) specifies the
baseline scale of uncertainty for each group of perturbed parameters. We
use
\[
\sigma_\lambda = 0.15,\qquad
\sigma_\mu = 0.10,\qquad
\sigma_{\mathrm{NH}} = 0.20,
\]
where \(\sigma_\lambda\) is used for the lung cancer onset probability
\(\lambda_t\), \(\sigma_\mu\) is used for the competing-mortality
probability \(\mu_t\), and \(\sigma_{\mathrm{NH}}\) is used for the
natural-history transition probabilities \(p_{23}\), \(p_{26}\), and
\(p_{36}\). These baseline scales are guided by uncertainty ranges and
parameter groupings considered in prior lung cancer screening and
natural-history modeling studies~\citep{Toumazis2021RiskBasedFramework,Bach2003}.

The term
\(-\frac{1}{2}\alpha^2\sigma_\theta^2\) in
\eqref{eq:log_norm_perturbation} is a mean-correction term. It ensures
\(
\mathbb{E}\!\left[\rho_\theta^{(i)}\right]=1,
\)
so that the perturbations are centered around the nominal parameter
values. Moreover, since \(\xi_\theta^{(i)}\sim\mathcal{N}(0,1)\), the
multiplier \(\rho_\theta^{(i)}\) is log-normal with variance
\(
\mathrm{Var}\!\left(\rho_\theta^{(i)}\right)
=
\exp\!\left(\alpha^2\sigma_\theta^2\right)-1.
\)
Therefore, for a fixed baseline scale \(\sigma_\theta\), increasing
\(\alpha\) increases the dispersion of the perturbation multipliers without changing their mean. 
Thus, \(\alpha\) determines the perturbation intensity: small values of \(\alpha\) generate perturbations close to the nominal cancer-progression model, whereas larger values generate more dispersed perturbed transition kernels.

After the selected parameters are perturbed, the affected rows of the cancer-state transition matrix are normalized when needed so that each row remains a valid probability distribution. The smoking-status transition probability \(\widehat Q_t\), the observation probability \(\widehat Z_t\), and the absorbing-state rows are kept fixed at their nominal values. This procedure generates
\(M\) perturbed cancer-state transition kernels
\(
\widehat P_t^{(1)},\ldots,\widehat P_t^{(M)}.
\)
The same perturbed kernels are used to evaluate both the robust policy and
the nominal ENGAGE policy, so that the comparison is based on identical
out-of-sample environments.

\subsection{Sensitivity of the male-cohort LCD and FP pattern at
\(\epsilon=0.10\) to random seeds}
\label{app:sanity_check}

In the main LCD and FP analysis for male heavy smokers, the smallest
ambiguity radius \(\epsilon=0.10\) produces a pattern that differs from
the results observed at larger radii. Specifically, at
\(\epsilon=0.10\), the robust policy yields negative
\(\Delta\mathrm{FP}\) and positive \(\Delta\mathrm{LCD}\) across the
three perturbation intensities
\(\alpha \in \{0.5,1.0,3.0\}\). In contrast, at moderate and large
ambiguity radii, the robust policy typically yields negative
\(\Delta\mathrm{LCD}\) and positive \(\Delta\mathrm{FP}\).

To examine whether the small-radius pattern is driven by  random sampling variation in the perturbation and simulation procedures, we repeat the perturbation evaluation under three independent random seeds, \(\{20240101, 12345, 987654321\}.\)
For each seed, we use the same simulation configuration as in the main analysis of Section \ref{sec:lcd_fp}, with \(M=300\) perturbation samples and \(N=100{,}000\)
individuals per sample. We evaluate the smallest radius
\(\epsilon=0.10\), together with two comparison radii,
\(\epsilon=0.90\) and \(\epsilon=1.70\), across the three perturbation intensities. The results are rounded to the nearest integer and reported in
Table~\ref{tab:sanity_seeds}.

\begin{table}[h]
\centering
\caption{Sensitivity of the male-cohort LCD and FP to random seeds.
\label{tab:sanity_seeds}}

\footnotesize
\begin{tabular}{cccc}
\toprule
$\alpha$ & $\epsilon$ & $\Delta\text{LCD}\ [\text{95\%\ CI}]$ & 
$\Delta\text{FP}\ [\text{95\%\ CI}]$ \\
\midrule
\multicolumn{4}{l}{Seed: 20240101}\\
0.5 & 0.10 & $+9\ [+1,\,+17]^{**}$  & $-1973\ [-2354,\,-1592]$ \\
    & 0.90 & $-10\ [-18,\,-2]^{**}$ & $+5667\ [+5287,\,+6046]$ \\
    & 1.70 & $-12\ [-20,\,-3]^{**}$ & $+5667\ [+5282,\,+6052]$ \\
1.0 & 0.10 & $+7\ [-2,\,+15]^{\,\text{FP}^*}$  & $-2612\ [-3027,\,-2196]$ \\
    & 0.90 & $-16\ [-23,\,-8]^{**}$ & $+4956\ [+4539,\,+5373]$ \\
    & 1.70 & $-10\ [-18,\,-2]^{**}$  & $+4080\ [+3642,\,+4519]$ \\
3.0 & 0.10 & $+8\ [-1,\,+16]^{\,\text{FP}^*}$  & $-3439\ [-3973,\,-2906]$ \\
    & 0.90 & $-31\ [-40,\,-22]^{**}$ & $+5427\ [+4774,\,+6080]$ \\
    & 1.70 & $-26\ [-36,\,-16]^{**}$ & $+2501\ [+1937,\,+3066]$ \\
\midrule
\multicolumn{4}{l}{Seed: 12345} \\
0.5 & 0.10 & $+7\ [-1,\,+15]^{\,\text{FP}^*}$ & $-2174\ [-2573,\,-1775]$ \\
    & 0.90 & $-19\ [-27,\,-11]^{**}$ & $+5463\ [+5063,\,+5863]$ \\
    & 1.70 & $-17\ [-25,\,-9]^{**}$ & $+5368\ [+4967,\,+5769]$ \\
1.0 & 0.10 & $+11\ [+3,\,+20]^{**}$ & $-2873\ [-3298,\,-2447]$ \\
    & 0.90 & $-12\ [-20,\,-4]^{**}$ & $+4784\ [+4348,\,+5219]$ \\
    & 1.70 & $-9\ [-17,\,-1]^{**}$  & $+3931\ [+3496,\,+4366]$ \\
3.0 & 0.10 & $+10\ [+1,\,+19]^{**}$ & $-3468\ [-3985,\,-2952]$ \\
    & 0.90 & $-27\ [-39,\,-16]^{**}$ & $+5074\ [+4448,\,+5701]$ \\
    & 1.70 & $-31\ [-43,\,-19]^{**}$ & $+2565\ [+2042,\,+3087]$ \\
\midrule
\multicolumn{4}{l}{Seed: 987654321} \\
0.5 & 0.10 & $+7\ [-1,\,+15]^{\,\text{FP}^*}$ & $-2121\ [-2523,\,-1720]$ \\
    & 0.90 & $-14\ [-22,\,-6]^{**}$ & $+5514\ [+5112,\,+5916]$ \\
    & 1.70 & $-16\ [-24,\,-8]^{**}$ & $+5403\ [+5001,\,+5805]$ \\
1.0 & 0.10 & $+15\ [+7,\,+23]^{**}$ & $-2844\ [-3266,\,-2422]$ \\
    & 0.90 & $-17\ [-25,\,-9]^{**}$ & $+4740\ [+4315,\,+5164]$ \\
    & 1.70 & $-17\ [-25,\,-9]^{**}$ & $+4086\ [+3645,\,+4527]$ \\
3.0 & 0.10 & $+4\ [-4,\,+12]^{\,\text{FP}^*}$ & $-3054\ [-3579,\,-2529]$ \\
    & 0.90 & $-25\ [-35,\,-16]^{**}$ & $+5291\ [+4674,\,+5908]$ \\
    & 1.70 & $-19\ [-28,\,-9]^{**}$ & $+2510\ [+1984,\,+3035]$ \\
\bottomrule
\end{tabular}

\vspace{0.8em}

\begin{minipage}{\linewidth}
\footnotesize
The table reports paired differences between the robust policy and the
nominal ENGAGE policy for male heavy smokers under three independent
random seeds. Each setting uses \(M=300\) perturbation samples and
\(N=100{,}000\) simulated individuals per sample. ``\(^{**}\)'' denotes
that both \(\Delta\mathrm{LCD}\) and \(\Delta\mathrm{FP}\) have 95\%
confidence intervals excluding zero; ``FP\(^{*}\)'' denotes that only
\(\Delta\mathrm{FP}\) is statistically significant.
\end{minipage}

\end{table}

Table~\ref{tab:sanity_seeds} shows that the direction of the
male-cohort LCD and FP pattern is reproducible across random seeds. At the
smallest ambiguity radius \(\epsilon=0.10\), all combinations of random
seed and perturbation intensity yield
\(\Delta\mathrm{LCD}>0\) and \(\Delta\mathrm{FP}<0\). This is the same
directional pattern observed in the main analysis: relative to the
nominal ENGAGE policy, the robust policy produces more LCDs but fewer
FPs at \(\epsilon=0.10\).

By contrast, at the comparison radii \(\epsilon=0.90\) and
\(\epsilon=1.70\), all combinations of random seed and perturbation
intensity yield \(\Delta\mathrm{LCD}<0\) and
\(\Delta\mathrm{FP}>0\). This matches the dominant pattern reported in
the main analysis, where the robust policy reduces LCDs relative to
ENGAGE but produces additional FPs at moderate and large ambiguity
radii.

These results indicate that the distinct LCD and FP pattern for male heavy
smokers at \(\epsilon=0.10\) is not specific to a single random seed.
However, the statistical significance of the positive
\(\Delta\mathrm{LCD}\) component is not uniform across seeds, which is
consistent with the small magnitude of this effect. We therefore
interpret the \(\epsilon=0.10\) result as a localized small-radius outcome pattern rather than as the dominant LCD and FP pattern of the robust
POMDP.

\subsection{Screening Efficiency Metric}
\label{app: Screening Efficiency Metric}

This appendix describes the screening efficiency metric used in the
sex-specific analysis in Section~\ref{sec:lcd_fp}. The metric measures the
number of additional false positives incurred by the robust policy per lung
cancer death averted, relative to the ENGAGE baseline, when both policies
are evaluated under identical perturbation samples.

For each perturbation intensity \(\alpha\) and ambiguity radius
\(\epsilon\) reported in Tables~\ref{tab:lcdfp_female_seer05} and
\ref{tab:lcdfp_male_seer05}, we compute the paired differences
\(\Delta\mathrm{LCD}\) and \(\Delta\mathrm{FP}\) as the sample means of the
robust-policy outcomes minus the ENGAGE-policy outcomes across
\(M=300\) perturbation samples. Each perturbation sample simulates
\(N=100{,}000\) individuals. Thus, \(\Delta\mathrm{LCD}<0\) indicates that
the robust policy produces fewer lung cancer deaths than ENGAGE, whereas
\(\Delta\mathrm{FP}>0\) indicates that it produces more false positives.

For qualifying cells, we compute the screening efficiency ratio as
\(
\frac{\Delta\mathrm{FP}}{|\Delta\mathrm{LCD}|}.
\)
A qualifying cell is one in which \(\Delta\mathrm{LCD}<0\),
\(\Delta\mathrm{FP}>0\), and the 95\% confidence intervals for both paired
differences exclude zero. The ratio represents the additional
false-positive burden per lung cancer death averted by the robust policy
relative to ENGAGE. Smaller values therefore indicate higher screening
efficiency.

These restrictions ensure that the ratio is interpreted only in settings
where the robust policy exhibits a statistically reliable mortality benefit
and where the associated increase in false positives is also statistically
reliable. Cells in which the robust policy does not reduce lung cancer
deaths are excluded because the denominator would not represent averted
deaths. Cells in which the robust policy does not increase false positives
are also excluded because the ratio would no longer measure the additional
false-positive burden associated with the mortality reduction.

Table~\ref{tab:efficiency} reports the screening efficiency ratio for all qualifying cells, together with sex-stratified summary statistics. These
values support the sex-specific efficiency comparison discussed in Section~\ref{sec:lcd_fp}.

\begin{table}[htbp]
\centering
\caption{Screening efficiency $\Delta\text{FP}/|\Delta\text{LCD}|$ of 
the Robust POMDP relative to the ENGAGE baseline among heavy smokers, 
restricted to cells in which both $\Delta\text{LCD}$ and $\Delta\text{FP}$ 
are significant at the $**$ level and $\Delta\text{LCD} < 0$. 
\label{tab:efficiency}}

\begin{tabular}{cccrr}
\toprule
$\alpha$ & $\epsilon$ & $\Delta\text{LCD}$ & $\Delta\text{FP}$ & 
$\Delta\text{FP}/|\Delta\text{LCD}|$ \\
\midrule
\multicolumn{5}{l}{Female} \\
1.0 & 0.90 & $-10$ & $+1{,}231$ & 123.1 \\
    & 1.70 & $-13$ & $+1{,}254$ &  96.5 \\
    & 1.90 & $-16$ & $+1{,}240$ &  77.5 \\
\addlinespace
3.0 & 0.50 & $-28$ & $+3{,}241$ & 115.8 \\
    & 0.90 & $-32$ & $+3{,}956$ & 123.6 \\
    & 1.70 & $-38$ & $+3{,}966$ & 104.4 \\
    & 1.90 & $-36$ & $+3{,}966$ & 110.2 \\
\midrule
\multicolumn{5}{l}{\textit{Female summary ($n = 7$): mean $107.3$, 
range $77.5$--$123.6$}} \\
\midrule
\multicolumn{5}{l}{Male} \\
0.5 & 0.90 & $-10$ & $+5{,}667$ & 566.7 \\
    & 1.70 & $-12$ & $+5{,}667$ & 472.2 \\
    & 1.90 & $ -9$ & $+5{,}582$ & 620.2 \\
\addlinespace
1.0 & 0.50 & $-15$ & $+4{,}931$ & 328.7 \\
    & 0.90 & $-16$ & $+4{,}956$ & 309.8 \\
    & 1.70 & $-10$ & $+4{,}080$ & 408.0 \\
    & 1.90 & $-12$ & $+3{,}896$ & 324.7 \\
\addlinespace
3.0 & 0.50 & $-23$ & $+4{,}428$ & 192.5 \\
    & 0.90 & $-31$ & $+5{,}427$ & 175.1 \\
    & 1.70 & $-26$ & $+2{,}501$ &  96.2 \\
    & 1.90 & $-26$ & $+2{,}414$ &  92.8 \\
\midrule
\multicolumn{5}{l}{\textit{Male summary ($n = 11$): mean $326.1$, 
range $92.8$--$620.2$}} \\
\bottomrule
\end{tabular}

\vspace{0.8em}

\begin{minipage}{\linewidth}
\footnotesize
Smaller 
values indicate higher screening efficiency. Summary statistics are 
reported as mean over qualifying cells.
\end{minipage}

\end{table}

\end{document}